\PassOptionsToPackage{table}{xcolor}
\documentclass[sigconf, authorversion, nonacm, balance=false]{acmart}
\usepackage{popets}

\usepackage[htt]{hyphenat}
\usepackage{listings}
\usepackage{wasysym}
\usepackage{multirow}
\usepackage{threeparttable}
\usepackage{makecell}
\newenvironment{boxlabel}
  {\begin{list}
    {$\bullet$}
    {
     \setlength{\labelwidth}{3em}
     \setlength{\labelsep}{0.5em}
     \setlength{\itemsep}{2pt}
     \setlength{\leftmargin}{0.5cm}
     \setlength{\rightmargin}{1em}
     \setlength{\itemindent}{0em} 
    }
  }
{\end{list}}

\definecolor{codegreen}{rgb}{0,0.6,0}
\definecolor{codegray}{rgb}{0.5,0.5,0.5}
\definecolor{codepurple}{rgb}{0.58,0,0.82}
\definecolor{backcolour}{rgb}{0.95,0.95,0.92}

\lstdefinestyle{mystyle}{
    backgroundcolor=\color{backcolour},   
    commentstyle=\color{codegreen},
    keywordstyle=\color{magenta},
    numberstyle=\tiny\color{codegray},
    stringstyle=\color{codepurple},
    basicstyle=\ttfamily\footnotesize,
    breakatwhitespace=false,         
    breaklines=true,                 
    captionpos=b,                    
    keepspaces=true,                 
showspaces=false,                
    showstringspaces=false,
    showtabs=false,                  
    tabsize=2
}

\setcopyright{popets}
\copyrightyear{YYYY}

\acmYear{YYYY}
\acmVolume{YYYY}
\acmNumber{X}
\acmDOI{XXXXXXX.XXXXXXX}
\acmISBN{}
\acmConference{Proceedings on Privacy Enhancing Technologies}
\begin{document}

\title[Your DRM Can Watch You Too]{Your DRM Can Watch You Too: Exploring the Privacy Implications of Browsers (mis)Implementations of Widevine EME}

\author{Gwendal Patat}
\affiliation{\institution{Univ Rennes, CNRS, IRISA}
  \city{}
  \country{}}
\email{gwendal.patat@irisa.fr}

\author{Mohamed Sabt}
\affiliation{\institution{Univ Rennes, CNRS, IRISA}
  \city{}
  \country{}}
\email{mohamed.sabt@irisa.fr}

\author{Pierre-Alain Fouque}
\affiliation{\institution{Univ Rennes, CNRS, IRISA}
  \city{}
  \country{}}
\email{pierre-alain.fouque@irisa.fr}

\renewcommand{\shortauthors}{Patat et al.}

\begin{abstract}
Thanks to HTML5, users can now view videos on Web browsers without installing plug-ins or relying on specific devices. In 2017, W3C published Encrypted Media Extensions (EME) as the first official Web standard for Digital Rights Management (DRM), with the overarching goal of allowing seamless integration of DRM systems on browsers. EME has prompted numerous voices of dissent with respect to the inadequate protection of users. Of particular interest, privacy concerns were articulated, especially that DRM systems inherently require uniquely identifying information on users' devices to control content distribution better. Despite this anecdotal evidence, we lack a comprehensive overview of how browsers have supported EME in practice and what privacy implications are caused by their implementations. In this paper, we fill this gap by investigating privacy leakage caused by EME relying on proprietary and closed-source DRM systems. We focus on Google Widevine because of its versatility and wide adoption. We conduct empirical experiments to show that browsers diverge when complying EME privacy guidelines, which might undermine users' privacy. For instance, we find that many browsers gladly give away the identifying Widevine Client ID with no or little explicit consent from users. Moreover, we characterize the privacy risks of users tracking when browsers miss applying EME guidelines regarding privacy. Because of being closed-source, our work involves reverse engineering to dissect the contents of EME messages as instantiated by Widevine. Finally, we implement EME Track, a tool that automatically exploits bad Widevine-based implementations to break privacy. 
\end{abstract}


\maketitle

\section{Introduction} \label{sec:introduction}

The use of the Web for streaming video services has increased tremendously in past years. According to~\cite{video_streaming_money}, the video streaming market is projected to grow from \$473 billion in 2022 to \$1,690 billion by 2029. In a business model based on subscription, video services protect their streams via encryption. These video services put the key in a ``DRM license''. If a user is subscribed, then the DRM license is fetched from the DRM server, and only the related DRM module, called CDM for Content Decryption Module, can get the key and decrypt the video. We call the pair (license server, CDM) a DRM key system. Multiple actors are involved in providing DRM key systems, three of which are distinguished: Microsoft PlayReady~\cite{playready}, Google Widevine~\cite{widevineSite} and Apple FairPlay~\cite{fairplay}. Naturally, different platforms support different DRM key systems. For instance, video streaming is protected by FairPlay on Apple devices (i.e., iOS devices, Apple TV, and Safari on macOS), while Edge on Windows can rely on PlayReady. The lack of cross-platform DRM compatibility was a major reason behind DRM being ultimately abandoned for iTunes Music in 2007.

DRM and the Web are no strangers since users often stream video on browsers. Historically, DRM on the Web was supported in plug-ins for a long time (e.g., Microsoft Silverlight and Adobe Flash). However, the advent of HTML5-based media playback systems encouraged the media industry to make DRM integration more seamless. In 2012, Google and Microsoft partnered with Netflix (a content provider) to propose a ``built-in'' DRM extension for the Web: the W3C Encrypted Media Extensions (EME)~\cite{EME}, with the overarching goal to define a standard DRM API (Application Programming Interface) that would work across multiple browsers or operating systems on a broad of range of devices. The EME specification does not create yet-another DRM key system. Instead, it allows browsers to discover, select and interact with any DRM module automatically. Thus, EME removes the burden of implementing content protection from browsers, whose role is now restricted to only redirecting DRM messages to the right native key system. In 2016, all major browser vendors demonstrated interoperable support of EME, one year before becoming a W3C Recommendation~\cite{eme_recommendations}. 

Standardizing EME received much controversy, as EME stands at the intersection between the desire of freedom of the Internet and the increasing need to protect premium contents and services. On the one hand, the EME project received the Emmy Award in 2019~\cite{eme_emmy_award}. Both the W3C Director (Tim Berners-Lee)~\cite{eme_debate_1} and the W3C CEO (Jeff Jaffe)~\cite{eme_debate_2} strongly defend EME adoption. On the other hand, opponents, such as Free Software Foundation, criticized W3C for standardizing DRM at the behest of a few actors in private industry and qualified the Firefox decision to support it as ``shocking'' and ``unfortunate''~\cite{eme_fsf_condemns}. As summarized in~\cite{eme_critics}, privacy received much attention among the voices of dissent.

\noindent
\textbf{Problem Statement.} 
Indeed, a DRM key system could cause user tracking because it might manipulate user-specific information that are either disclosed in EME messages or persistently stored on the user's device. The starting point of our work is that, as acknowledged by EME, ``\textit{privacy cannot be met without the knowledge of the privacy properties of the Key System and its implementations}''. Despite the much-raised controversy, little has been done to understand the privacy properties guaranteed by existing key systems, and how they are related to the EME ones. As EME is already widely deployed~\cite{eme_browsers}, we take a closer look at the privacy leakage of EME implementations in practice. In this paper, we conduct, to the best of our knowledge, the first privacy evaluation of EME, covering the format of its messages and the impact of browsers default configurations on user tracking. In particular, we explore the tension between the EME privacy guidelines and its actual compliant modules. We only focus on EME as implemented by Widevine for two main reasons. First, Widevine is the most versatile DRM key system, since it is deployed on various operating systems (e.g., Linux, Windows, macOS, and Android) on various devices, ranging from personal computers to smartphones, smart TVs, and embedded devices. The scope of other key systems, such as PlayReady and FairPlay, would have been more limited. Second, each key system introduces numerous proprietary technical details, and we preferred clarity in our presentation and longitudinal analysis over completeness, especially since our conclusion would have remained the same: there is a real risk of privacy leakage caused by EME key systems.
\\

\noindent
\textbf{Findings Summary.} 
Our main goal is to gain insights into how the opaque Widevine EME can lead to privacy leaks. In other words, we investigate whether users have any guarantee about their privacy with EME relying on proprietary and closed-source DRM modules like Widevine. We refine our goal into two high-level questions: (1) is there any gap between the EME privacy guidelines and the browser implementations of Widevine-based EME? and (2) what is the privacy impact of this gap? To answer our questions, we first reverse engineer the EME messages as defined by Widevine. Second, we examine the privacy mechanisms put in place by Widevine on Windows, Linux, and Android. We conduct our experiments with a selection of popular browsers as well as privacy-focused ones. Third, supported by our experimental setup and reverse engineering, we find that EME Widevine misses applying some important EME recommendations. In our analysis, we focus on the Widevine distinctive identifier, namely Client ID, and session-related stored data, namely persistent sessions. In that sense, we find that all desktop browsers fail to encrypt the Client ID containing identifying information about the user's device. Ironically, we find that some mobile privacy-focused browsers, such as Ghostery, do not encrypt the Client ID, which can be obtained by a few JavaScript calls with no or little explicit users consent, as desired by EME designers~\cite{w3c_archives}. In addition, we highlight some mobile browsers storing Widevine data as cookies even when first-party cookies are not enabled. Fourth, we continue our study and inspect Widevine device fingerprinting and resulting user tracking. Here, among others, we examine the nature of the Client ID and look deeper into its different fields, uniqueness, and stability. We show that this Client ID allows not only building an augmented User-Agent on which browsers have no control, but also a stable and unique fingerprint on Android mobile devices. Overall, our work shows empirical evidence that EME constitutes an effective privacy leakage vector, implying that the W3C claims about ``greater privacy'' guaranteed by EME are just all red herrings. \\

\noindent
\textbf{Threat Model.}
By design, DRM systems are bound to some individualization process. Indeed, despite EME recommendations, Widevine establishes a factory provisioning process to embed some unique Device ID on the client side. Widevine defines a privacy mode in which this unique identifier is encrypted before being sent to any remote server. Only Widevine and its partnered service providers, aka Over-The-Top (OTT) platform (e.g., Netflix, Disney+), can decrypt it, implying that they can effortlessly fingerprint all devices. Widevine does not publicly state the required terms for being a partner OTT. This could be concerning, but we consider a wider threat model: an attacker (website) with no Widevine partnership, meaning not having a Widevine-signed server certificate, that can call EME JavaScript API (see \autoref{subsec:permissions} for required permissions) acting as a mirrored proxy between a client and a license server. In this paper, we show that this attacker can achieve three goals: 
\begin{enumerate}
	\item Fingerprint users on Android devices if they are on a browser not applying the privacy mode (e.g., firefox).
	\item Get identifying, but not unique, attributes if the privacy mode is not always enforced (e.g., all desktop browsers). 
	\item Achieve surreptitious cross-session tracking on the same origin even when cookies are blocked and site data are wiped. For example, this allows an attacker to link user activity on the same site. 
\end{enumerate}

\noindent
\textbf{Contributions.}
We summarize our contributions as follows:
\begin{boxlabel}
	\item We reverse engineer the EME messages as defined by Widevine.
	\item We analyze the browser implementations of Widevine-based EME with respect to the EME privacy guidelines.
	\item We explore the amount of privacy leakage whenever browsers do not strictly comply with EME guidelines.
	\item We implement and open-source EME Track; a tool leveraging our findings about EME to demonstrate user tracking.
	\item Our results were acknowledged by the Mozilla Hall of Fame.
\end{boxlabel} 
\section{Background}

\subsection{DRM Systems} Digital Rights Management (DRM) systems aim at protecting digital media from piracy while enabling distribution and consumption by legitimate users. Medias consist of separate video/audio tracks and subtitles. The DRM ecosystem enforces given business rules to protect media using two main entities: a License Server and a Content Decryption Module (CDM). In practice, a Content Delivery Network (CDN) supplies encrypted content to a client, and a license server provides the necessary keys to decrypt such content. These keys are commonly referred to as licenses. Only the DRM module on the user device can retrieve these keys and decrypt the content, which makes it possible to control media consumption.

The CDM constitutes the DRM module performing sensitive operations, such as decryption and license requests. Every DRM scheme, known as Key System, provides its own CDM that includes proprietary mechanisms for license server communication and rules about license usage and renewal. 

\subsection{Google Widevine} \label{subsec:widevinedrm}Among DRM systems, Widevine~\cite{widevineSite} is a closed-source proprietary DRM technology purchased by Google in 2010. Its early version, Widevine Classic, was compatible with older Android versions (up to Android 5.1) but only supported the \textit{.wmv} format. Nowadays, the most recent version, Widevine Modular, or simply Widevine, supports various streaming protocols, including MPEG-DASH and CENC~\cite{iso_dash_2022, iso_cenc_2016} regularly used by OTT platforms such as Netflix and Disney+. Widevine is now the most deployed DRM by being present in Android devices, Android TVs, most web browsers, and many other devices (e.g., Chromecast, VR Headsets).

\subsection{W3C EME Standard} Due to the fragmented ecosystem of DRM solutions in the 2010s, the World Wide Web Consortium (W3C) defined the Encrypted Media Extensions (EME) standard to provide a standardized API enabling web applications to interact with browser-supported DRMs. EME is designed to make the same web application run on any user-agent regardless of the underlying DRM implementation. Despite being optional, EME is supported by major browsers: Edge, Firefox, Chrome, Safari, Opera, and their mobile counterparts~\cite{eme_browsers}.

\subsubsection{EME Components} 
The EME API brings together multiple entities. Here, we list the main ones used during the control flow.

\noindent
\textbf{User-Agent.} The EME implementation lies in the user-agent (e.g., a web browser) and is called through JavaScript to communicate with a CDM and distant servers using opaque messages. In the rest of the paper, we use the terms browser and user-agent interchangeably.

\noindent
\textbf{CDM.} The Content Decryption Module DRM system in the EME specification. It is responsible for license protection and usage enforcement to decrypt media content on the user's device. 

\noindent
\textbf{CDN.} The Content Delivery Network is a distant server providing encrypted media to the user device on demand. It also offers the necessary information for license requests and DRM configurations.

\noindent
\textbf{License Server.} License requests generated by the CDM are sent to the related license server of the given key system. This server manages licenses and policies for protected content.

\begin{figure}
	\centering
	\resizebox{\columnwidth}{!}{\includegraphics[scale=1.175]{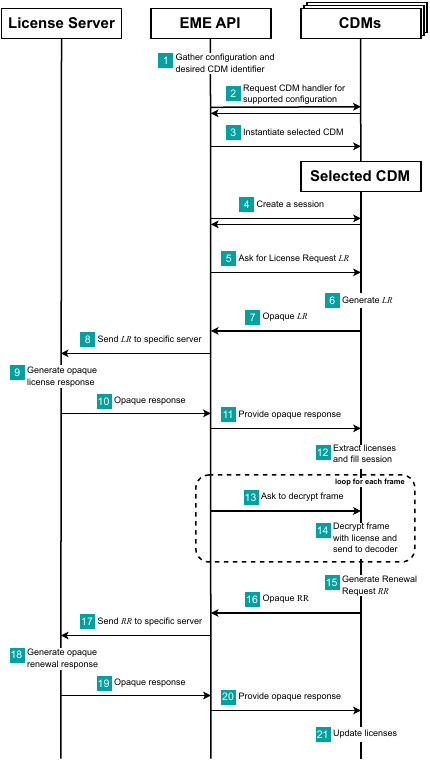}
	}
	\caption{EME Workflow: License Acquisition and Renewal.}
	\label{fig:eme_workflow}
	\vspace{-2em}
\end{figure} \subsubsection{Protocol Workflow} \label{subsub:eme_api}
EME defines a uniform API centered on CDM session management and license acquisition while abstracting message contents. Henceforth, the EME API relies on multiple objects and events to implement the proprietary protocols of CDMs. We exclude user authentication to license server and CDN from EME workflow due to it being out of scope of the EME specification.

\noindent
\textbf{License Acquisition.} An overview of the protocol can be seen in~\autoref{fig:eme_workflow}. For the execution, we assume that the EME user-agent has already received all the required information about the encrypted media from the CDN, such as supported codecs and used DRM system. The EME workflow starts in step~\framebox(7,7){\small 1} by the JavaScript running in the user-agent, taking as arguments the requested DRM as well as configurations for media decryption and playback. Using the self-assigned UUID (universally unique identifier)~\cite{dash_drm_id} of the chosen DRM system, the user-agent asks for a handler using the \texttt{requestMediaKeySystemAccess} method of the \texttt{navigator} object~\framebox(7,7){\small 2}. The EME API \texttt{createMediaKeys} is then used to instantiate the desired CDM in step~\framebox(7,7){\small 3}. At this stage, the CDM is available and represented by a \texttt{MediaKeys} object but cannot receive any license. Step~\framebox(7,7){\small 4} creates a session within the CDM, represented by a \texttt{MediaKeySession} object, and identified by a session ID. Such sessions refer to key wallets and cryptographic environments isolated from each other, used to receive licenses and decrypt protected media. On creation, a session key wallet is empty and cannot decrypt anything. To receive media decryption keys, referred to as licenses, media-specific \textit{Initialization Data} are sent to the CDM~\framebox(7,7){\small 5} to generate a license request using the \texttt{generateRequest} method. The content of this generated request fully depends on the key system in use~\framebox(7,7){\small 6} and is therefore managed as an opaque message by EME. When receiving the license request~\framebox(7,7){\small 7}, the user-agent forwards it to a CDM-compatible license server~\framebox(7,7){\small 8}. The opaque request is processed by the license server generating an opaque license response in step~\framebox(7,7){\small 9} to send back to the user-agent~\framebox(8,8){\small 10}. Using \texttt{update}~\framebox(8,8){\small 11}, the response is sent to the CDM session and proceeded by DRM-specific mechanisms~\framebox(8,8){\small 12}. Now containing at least one license, the CDM session can receive frames of the encrypted media to decrypt them~\framebox(8,8){\small 13}, and forward them in clear to the decoder~\framebox(8,8){\small 14}. At this point, the session can be terminated, or renewal can happen.

\noindent
\textbf{License Renewal.} Once received, licenses can be extended by renewal mechanisms. Based on DRM-specific license policies, the CDM can be authorized to ask for key updates. After policy verification, the CDM generates a renewal request~\framebox(8,8){\small 15} and provides it to the user-agent by the \texttt{MediaKeyMessageEvent} event~\framebox(8,8){\small 16}. Steps~\framebox(8,8){\small 17} to~\framebox(8,8){\small 19} are similar to the ones in the steps~\framebox(7,7){\small 8} to~\framebox(8,8){\small 10}. The renewal response is sent to the CDM with the \texttt{update} method~\framebox(8,8){\small 20} to refresh session licenses~\framebox(8,8){\small 21}.

\noindent
\textbf{Persistent Session.} In addition to license request/response API, EME also defines CDM session types: `\texttt{persistent-license}' and `\texttt{temporary}' types. Standard \texttt{MediaKeySession} are temporary sessions, meaning licenses will be destroyed on session closing. For persistent sessions, the user-agent and CDM need to support offline license storage. If license policies allow a persistent state, licenses can be stored on closing for future usage (e.g., offline streaming). They can therefore be reloaded using their session ID with the \texttt{load} function of a \texttt{MediaKeySession} object to fill the newly created session in step~\framebox(7,7){\small 4} of~\autoref{fig:eme_workflow}. 

\subsubsection{EME Privacy Concerns} \label{subsub:eme_privacy_concern} While non-normative, the EME standard raises multiple privacy concerns, especially regarding fingerprinting, information leakage, and user tracking. The raised concerns aim to guide the DRM providers while specifying the content of the different opaque messages between license servers and CDM modules. Among these issues, the potential presence of so-called \textit{Distinctive Identifier} and \textit{Distinctive Permanent Identifier} is highlighted multiple times since such information could lead to user tracking. As their names suggest, these identifiers are defined as unique or shared with a small number of users that could be used to identify them. The main difference between the two is the provenance of data. Distinctive Permanent Identifiers are any distinctive value associated with or derived from data being at least non-trivial for the user to remove, reset or change, such as hardware identifier, operating system, user-agent instances, or factory values. As for Distinctive Identifiers, they represent data produced or derived from one or more Distinctive Permanent Identifiers and exposed to a component other than the CDM but with the possibility of being cleared by the OS. As mentioned in the standard, due to their characterizing nature, such identifiers could be used by malicious origins for fingerprinting on user devices and user tracking.

Within the EME standard, user tracking is further discussed in the persistent session mechanism. As explained in section~\ref{subsub:eme_api}, this feature allows the CDM to store licenses for future usage. The specification states that the user-agent must allow the user to clear the stored licenses as site data, such as cookies.

To limit the impact of these possible issues, EME puts forward a per-origin and per-browsing profile policy for DRM system usage and data storage. Moreover, users should be asked for consent when a CDM is needed, and this consent should be linked to a specific origin without giving away permission to untrusted websites.

\subsection{Web Tracking} 

In practice, web user tracking can be achieved by different means categorized into two major classes: stateful and stateless tracking.

\subsubsection{Stateful Tracking}
Stateful tracking refers to any mechanism storing data on user devices to identify and track them in the future. The first example of such mechanisms are cookies, which have been first introduced to enable a stateful web experience but have been rapidly used to impact privacy by third-party websites~\cite{schwartz_2001}. Stored values could then be used across origins to identify users and trace navigation. With this first step, trackers pushed HTTP cookies further using similar storage components like Flash cookies~\cite{Ayenson2011FlashCA}, JavaScript properties, or even HTML5-specific mechanisms~\cite{10.5555/2038772.2038791}. By combining these systems, Cookie respawning and Evercookies~\cite{evercookie_samykamkar} were introduced to restore previously removed cookies. Using, for instance, Flash cookie, and HTML Storage API, multiple studies have shown its effectiveness in circumventing anti-trackers~\cite{DBLP:conf/aaaiss/SoltaniCMTH10, DBLP:conf/ccs/AcarEEJND14}.

\subsubsection{Stateless Tracking} 
In 2009, Mayer~\cite{mayer2009any} looked into whether the variations resulting from browser attributes could result in the deanonymization of web clients. He checked to see if a distant server could make user identification possible using variations in browsing contexts. He observed that a browser might display values or behaviors linked to the hardware, operating system, and browser setup creating a device fingerprint and enabling user tracking without the need for data to be stored on the user device. Since then, multiple studies have been made on even larger scales to inspect the impact of fingerprinting on web privacy~\cite{DBLP:conf/pet/Eckersley10, DBLP:conf/sp/LaperdrixRB16, DBLP:conf/www/Gomez-BoixLB18}. These studies have added different attributes to the fingerprint, such as fonts, timezones, JavaScript engine behaviors~\cite{MBYS11, mulazzani2013fast} or even specific APIs like WebGL or Canvas~\cite{MS12}, to generate stronger fingerprints both in terms of uniqueness and stability over time.

\subsubsection{Countermeasures} To circumvent both stateful and stateless tracking, defenses have been put in place to increase user privacy. User tracking through third-party cookies is one of the most used techniques for stateful tracking nowadays. In response, web browsers are now, for most, blocking third-party from using cookies by default. In addition, Google launched in 2022 the Privacy Sandbox~\cite{google_privacysandbox} initiative to produce new privacy-oriented web standards to remove third-party cookies and increase privacy boundaries across websites.

With stateless fingerprinting, two opposed trends can be observed; either increasing the fingerprint homogeneity among all users or breaking its stability. The first solution is the one chosen by the Tor browser~\cite{tor_browser} with the aim to blend users into the mass without distinctive attributes. Apple also went this way with the Safari web browser, using it to reduce differences between users~\cite{safari_fp}. Oppositely, web extensions have been developing to break fingerprinting stability over time, such as Canvas Defender~\cite{canvasdefender}, adding noise to the Canvas API calls between browsing sessions.

Another defense chosen by developers is to reduce the browser APIs surface to minimize the information that can be retrieved by origin. Web extensions, such as Canvas Blocker~\cite{canvasblocker} or NoScript~\cite{noscript}, have been proposed with these goals in mind by reducing APIs for untrusted websites or deactivating JavaScript support to avoid potential fingerprinting. Browsers have also incorporated built-in protection against fingerprinting like Tor or Brave~\cite{brave_browser} blocking by default several APIs or JavaScript execution. Firefox has put in place a by default Enhanced Tracking Protection~\cite{firefox_protection} blocking known trackers and fingerprint scripts based on a denylist to avoid negatively impacting users' browsing experience. 
\section{EME Widevine}
\label{sec:eme_wv}

As presented in~\autoref{subsec:widevinedrm}, Widevine is the most deployed DRM system in the wild and is, therefore, heavily used by web-based content providers through the EME standard interface. In this setting, Widevine specifies the content of the opaque messages in EME for both license acquisition and renewal. Related works dissect the Widevine CDM in Android environments~\cite{DBLP:conf/dsn/PatatSF22, DBLP:conf/sp/PatatSF22} and desktop ones~\cite{tomer_widevinel3decryptor_wiki, pywidevine}. In this section, we reverse engineer the EME instantiation by Widevine regarding license acquisition and persistent license. Our goal is not to explore all the underlying cryptographic operations but to study the content of the opaque messages. In addition, we examine the mechanisms put in place by Widevine to address the various EME privacy concerns. 

\subsection{Reverse Engineering Settings}
\label{subsec:re_settings}
We conducted a two-stage approach to inspect Widevine-based EME. The first settings are as follows: we leveraged some web extensions that monitor all calls to EME API. Then, we requested media from various content providers, and noticed that EME works similarly for both paid services (e.g., Netflix and Disney+) and experimental ones (e.g., Bitmovin). Finally, we examined all the collected messages to determine which parts are Widevine-defined and which are content provider-defined

As for the second settings, we developed a script calling EME APIs, and hosted it on our web server. Our script implements the workflow in \autoref{fig:eme_workflow} with Widevine as the underlying DRM system. To this end, we leveraged the integration platform made available by Widevine for external developers to perform license acquisition/renewal~\cite{widevine_integration_platform}. Finally, we accessed our script, while varying operating systems (i.e., Android, Windows, Linux, and macOS) and browsers (e.g., Firefox, Chrome, Edge). This allows us to fine-tune our understanding of different fields that keep unchanged or mutate for a given execution environment. 

\begin{table}[]
	\resizebox{\columnwidth}{!}{\begin{threeparttable}
		\begin{tabular}{lll}
			\hline
			\hline
			\textit{Widevine Protocol} & \textit{EME API}              & \textit{Message Content}               \\
			\hline
			\rowcolor{gray!20}
			License Request            & \texttt{generateRequest}      & \begin{tabular}[c]{@{}l@{}}Request ID  \\ Content Key ID(s)\\ Client ID\end{tabular} \\
			License Response           & \texttt{update}               & \begin{tabular}[c]{@{}l@{}} Request ID \\ Content Key(s)\\ TTLs\\ License Policy\end{tabular} \\
			\rowcolor{gray!20}
			Renewal Request            & \texttt{MediaKeyMessageEvent} & \begin{tabular}[c]{@{}l@{}}Request ID  \\ Client ID$^\dagger$ \\ License Policy\end{tabular} \\
			Renewal Response           & \texttt{update}               & \begin{tabular}[c]{@{}l@{}}Request ID  \\ Updated TTLs\\ License Policy\end{tabular} \\ \hline \hline
		\end{tabular}
		\begin{tablenotes}
			\small
			\item[$\dagger$] Optional, presence defined by license policies.\\
		\end{tablenotes}
	\end{threeparttable}
	}
	\caption{Content of Widevine Messages for License Acquisition and Renewal.}
	\label{table:content_messages}
	\vspace{-2em}
\end{table}

 \subsection{Opaque Messages in Widevine Workflow} \label{subsec:license_acquisition_widevine} As studied in~\cite{DBLP:conf/sp/PatatSF22}, the Widevine protocol defines four main operations: License/Renewal Requests and License/Renewal Responses. Here, we investigate the correspondence of these operations with the EME protocol. To acquire a media license, the EME user-agent creates a session within the Widevine CDM to ask for a license request through the \texttt{generateRequest} EME API. The CDM requires some \textit{Initialization Data} to issue licenses, also stated as \textit{Content Keys}. These data include Content Key IDs associated with some encrypted media. The corresponding request contains a Request ID for associating subsequent responses, one or several Content Key IDs for the desired media, and a Client ID used for license protection (further details are given in~\autoref{subsec:widevine_client_id}). Using the EME JavaScript API, the browser sends this request alongside its signature to a license server, which will respond with the according license response. This response contains the Content Keys for media decryption, alongside key usage control data: Time-To-Live (TTL) and license policies (refer to~\autoref{subsec:widevine_license_policy}). Using \texttt{update}, the response is then forwarded to the CDM session, which is then ready to decrypt the content.

During their lifetime, licenses can be renewed to extend license TTLs. The CDM can generate a renewal request if the license policy authorizes an extension to the expiration date. In Widevine, this message includes the Request ID of the previous license request, the related license policy, and, depending on this policy, the Client ID of the device. The user-agent receives this request as an EME event through the \texttt{MediaKeyMessageEvent} to send it to the corresponding server. The license server responds with an updated TTL, new license policy, and Request ID, all forwarded to the CDM using the \texttt{update} call like a usual license response. A summary of the content of opaque Widevine messages can be seen in~\autoref{table:content_messages}.

\subsection{Persistent Session} \label{subsec:license_storage_widevine} 
As explained in section~\ref{subsub:eme_api}, EME can define a persistent session type. Once closed, a persistent session can be re-opened to retrieve content keys without the need for a license acquisition phase. In Widevine, the workflow does not differ for a temporary or persistent session; the type is only mentioned when opening the session. However, a persistent session becomes persistent only when a response is processed through the \texttt{update} call and the associated license policies permit it. Upon session closing, Widevine entirely relies on the user-agent implementation to store a session blob from the CDM. Here the user-agent is responsible for implementing a per-origin policy as mentioned in EME using, for instance, service storage for licenses inside dedicated databases. Sessions can then be re-opened using their session ID through the \texttt{load} function.

\begin{table}[]
	\resizebox{\columnwidth}{!}{\begin{tabular}{lll}
	\hline
	\hline
	\multicolumn{1}{l}{\textit{Attribute}} & \multicolumn{1}{l}{\textit{Description}}  &  \multicolumn{1}{l}{\textit{Example}} \\ 
	\rowcolor{gray!20}

	\hline
	Architecture & CPU architecture & arm64-v8a\\
	Company Name & Device manufacturer & Google\\
	\rowcolor{gray!20}
	Device Name & Device codename & panther \\
	Product Name & Product codename & panther \\
	\rowcolor{gray!20}
	Model Name & Device model name & Pixel 7\\
	Platform Name & Desktop OS name & N/A\\
	\rowcolor{gray!20}
	Application Name & Calling Application & org.mozilla.firefox \\
	Package Cert Hash & \Gape[0pt][2pt]{\makecell[l]{Hash of\\Application Certificate}} & p4ti...xmwQ= \\
	\rowcolor{gray!20}
	Build Info & Complete build name & \Gape[0pt][2pt]{\makecell[l]{google/panther...\\8940162..keys}}\\
	CDM Version &  \Gape[0pt][2pt]{\makecell[l]{CDM protocol\\version number}} & 17.0.0\\
	\rowcolor{gray!20}
	\Gape[0pt][2pt]{\makecell[l]{Security Patch Level\\(SPL)}} & OEM Crypto SPL & 0\\
	OEM Build Info & \Gape[0pt][2pt]{\makecell[l]{OEM Crypto level\\and build date}} & \Gape[0pt][2pt]{\makecell[l]{OEMCrypto Level3\\Code May 20 2022\\21:36:542}} \\
	\hline
	\hline
	\end{tabular}
	}
	\caption{Widevine Client Info.}
	\label{tab:widevine_client_info}
	\vspace{-3em}
\end{table} \subsection{Widevine License Policy} \label{subsec:widevine_license_policy} 
License policies are included in several Widevine EME messages. Their goal is to add management rules over content licenses. These rules are defined on the license server side and are enforced by the CDM. A license server can define such policy to control renewal and persistent rights, the timing for renewal requests, license and media playback duration, or rental duration for offline licenses. Among these settings, a license server can ask the CDM to always include the Client ID within license renewal requests. By leveraging the Widevine Integration Platform, allowing a set of policies to be defined by developers for integration testing, we reversed the inline policies sent to the license server to identify various policy settings. The resulting correlation between inline codes and policies can be seen in~\autoref{appendix:widevine_license_policy}; note that these codes might not be exhaustive.

\subsection{Widevine Client ID} \label{subsec:widevine_client_id} 
In Widevine, the license request includes a Client ID being 40 times the size of the request ID and key ID combined, representing almost the entire size of the request. The Client ID consists of the Client Info and the Device RSA Public Key. These Client Info contain various metadata providing information about device-specific software versions and architectures. Among them, we can find the CPU architecture, OS name, Original Equipment Manufacturer (OEM) Crypto library version number, model name, or the complete device build name. The complete list of Client Info attributes can be seen in~\autoref{tab:widevine_client_info}. Note that some attributes do not apply to desktop implementations; for instance, desktop Client Info contain the architecture, company, model, platform, and CDM version.

\begin{figure}
	\centering
	\resizebox{\columnwidth}{!}{\includegraphics[scale=1]{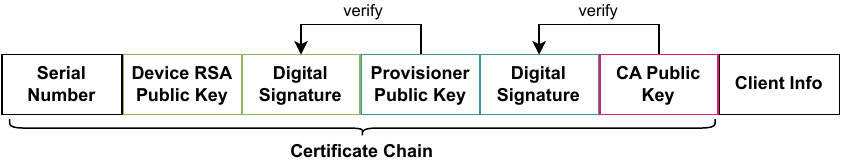}
	}
	\caption{Client ID Fields.}
	\label{fig:client_id}
	\vspace{-1.5em}
\end{figure}

In addition to metadata, the Client ID includes a DRM certificate chain composed of the device's DRM public key, called the \textit{Device RSA Public Key}, signed up to Widevine root certificate. On request, the Device RSA Public Key is used by the license server to encrypt licenses based on a Widevine-specific crypto key ladder~\cite{DBLP:conf/sp/PatatSF22}, while its corresponding private key is used by the CDM to sign license requests. The structure of the Client ID can be seen in~\autoref{fig:client_id}.

\subsection{Privacy Protection Mechanisms: Privacy Mode \& VMP} \label{subsec:privacy_vmp}
The inner structure of the Client ID used during the license acquisition of Widevine has brought potential privacy issues correlated to the ones raised by the EME standard. Indeed, the Client ID might constitute a \textit{Distinctive Permanent Identifier}, since it includes architecture, company, build, or even the certificate chain. As an example, \autoref{tab:widevine_client_info} shows the identifying characteristic of a Pixel 7. 

To avoid the leakage of such data, Widevine has put in place the \textit{Privacy Mode}. This mode needs to be enabled by the user-agent and forces the CDM to generate a \textit{privacy key} to encrypt the Client ID within requests between the license server and the CDM. A privacy key is generated for each new request and used to encrypt the Client ID using AES in CBC mode with random IV. This key, which is a 128-bit AES key, is then encrypted by a \textit{Server Certificate}, sometimes referred to as \textit{Service Certificate}, and placed within the request. This server certificate is either provided by the default CDM instance configuration or by the license server using the EME API \texttt{setServerCertificate}. When provided by a server, its authenticity is verified by a hard-coded public key within the CDM.

Closely related to the Privacy Mode, the Verified Media Path (VMP) is used to verify the integrity of the CDM using cryptographic signatures to enforce a protected video decryption and decoding chain. VMP implicitly mandates the usage of Privacy Mode in desktop implementations~\cite{widevine_oldnews}. VMP is supported in macOS and Windows, but not Linux. Android systems rely on hardware protection for secure video decoding. VMP is, therefore, not necessary, leaving the Privacy Mode activation to the user-agents.

\section{User-Agents Implementations of Privacy Mechanisms for Widevine}
\label{sec:ua_implem_privacy_mechanisms}

\subsection{Motivations and Research Questions}
Recall that the main goal of our study is to gain insights into how the opaque Widevine EME can lead to privacy leaks in practice. To this end, we aim to answer the following high-level questions:
\begin{enumerate}
\item Is there any gap between the recommendations in the EME specifications and the browser implementations?
\item What is the privacy impact of this gap?
\end{enumerate}

In this section, we begin by refining the first question above into three research questions (RQs) that guide our search for a better understanding of Widevine EME in the wild. The opaque nature of Widevine makes this task daunting since there is little documentation related to this topic. Thus, we develop a systematic approach to examine how browsers comply with the EME privacy guidelines for Widevine. We detail the second question in \autoref{sec:privacy_considerations}.

As mandated by EME, the usage of a DRM system shall not reveal any identifying information about the user device. Our reverse engineering of the Widevine EME (refer to \autoref{sec:eme_wv}) highlights the primary sources of privacy leakage in Widevine: (1) the Client ID that contains identifying or unique data per device, and (2) the data stored for persistent sessions, since they might behave as cookies. While Widevine has grown to be a multi-billion dollar industry and is basically available on every platform, the Widevine EME has never been subjected to a comprehensive privacy assessment. Accordingly, our study aims to shed light on the less-understood privacy impacts of Widevine in the context of EME implementations. We present our study that answers the following questions with a methodology involving various empirical evidence and observations. Indeed, we evaluated nine mobile browsers and six of their desktop counterparts, including both mainstream and privacy-focused ones. With our study, we aim to answer the following Research Questions that arise:
\begin{itemize}
\item \textbf{RQ1:} Do browsers activate the Privacy Mode of Widevine by default for license acquisitions?
\item \textbf{RQ2:} Does the Privacy Mode protect all occurrences of the Client ID in the Widevine workflow?
\item \textbf{RQ3:} How does the data related to Persistent Sessions behave like cookies? 
\end{itemize}

\subsection{Browser Selection}
To investigate the privacy mechanisms of Widevine on desktops and Android mobiles, we selected popular web browsers based on their market share, as well as privacy-focused ones according to the user base and download number. The exact version for each selected browser can be seen in~\autoref{appendix:browser_versions}.

On desktop, we opt for Chrome, Edge, Firefox, and Opera for mainstream browsers~\cite{browserDesktopMarketShare}. For privacy-oriented ones, we got Brave and Tor with respectively 50M and 2.5M monthly users~\cite{tor2022users, brave50Mactiveusers}.

On Android, we used the Android versions of Chrome, Edge, Firefox, and Opera, to which we added the Samsung Internet Browser to represent the majority of users~\cite{browserMobileMarketShare}. For privacy-focus ones, claiming enhanced features against user tracking and fingerprinting, we inspected Ghostery and Firefox Focus with 1M+ and 10M+ downloads on the Google Play Store~\cite{ghostery_gp, firefoxfocus_gp} respectively. In addition, we included the mobile apps of Brave and Tor.

Note that for this study, Safari is excluded since the Widevine DRM is not supported, and replaced when on macOS by Apple own key system FairPlay, being out-of-scope of this study. We do not consider iOS, since Widevine is not supported on it.

\subsection{Experimental Design} \label{subsec:implem_methodo}
\subsubsection{Browsers Setup} During our experimentation, we categorized desktop and Android mobile browsers regarding their underlying implementations into two main families: Chromium-based and Firefox-based. This distinction allowed us to establish similarities in the results. In addition, we tested desktop browsers both with VMP-compatible OS (macOS and Windows) and non-VMP, such as Linux systems, for Privacy Mode enforcement testing. Regarding browser configuration, we tested the default settings and privacy-enhancing features when available.

\subsubsection{RQ1}
The Privacy Mode is defined to protect the Client ID. Some might reasonably assume that the Widevine CDM enforces it by default regardless of the underlying browsers or operating systems. Thus, the Client ID is always encrypted, and never appears in clear while being transferred. To challenge this claim, we perform both static and dynamic analyses. First, we rely on our reverse engineering settings (refer to \autoref{subsec:re_settings}) to ask Widevine to start the license acquisition protocol and intercept the generated license request. Note that we keep the default configurations while making the request. We consider that Privacy Mode is not enabled by default if the obtained Client ID is not encrypted. Second, we statically inspect browsers' binaries and source code when available to search for any Widevine Privacy Mode related calls. In particular, Android allows explicit activation of the Privacy Mode through its DRM API. We look for these corresponding calls. To the best of our knowledge, there is no equivalent call on desktops.

\subsubsection{RQ2} 
The Client ID is present in the license request, which is encrypted when the Privacy Mode is enabled. Based on \autoref{table:content_messages}, the Client ID can also be present in the renewal request. Therefore, we answer our question by making Widevine generate license renewals to look into their content for the Client ID. Unlike license requests, EME does not provide a public API to create renewal requests. Indeed, anticipating the expiration of a key, the Widevine CDM triggers an event including the renewal request to be forwarded to the license server. The problem is that the Widevine test license server in our experimentation settings provides licenses that do not issue any renewal event. In order to overcome this limitation, we leveraged once again the Widevine Integration Platform. Recall that for testing purposes, this platform allows us to define license policies through an argument within the license server URL. Thus, we specify the policies so that a renewal request is generated a few seconds after loading the related license. We also mandate the optional Client ID to be included in this request. Finally, we automate our script to recover the license and renewal requests from the EME workflow that is supported by numerous browsers in various environments. This question evaluates the adequacy of Privacy Mode implementations in protecting the Client ID.

\subsubsection{RQ3}
To address this question, we need to open persistent sessions by the Widevine CDM through custom license policies allowing offline license storage. Here, we verified three properties related to cookies. First, we check whether persistent sessions can be opened even when first-party cookies are blocked. Second, we attempt to re-open the closed persistent sessions from different origins and browsing profiles. Third, we study the impact of browsers deleting cookies and site data on the stored data. On mobiles, we went further and deleted all the data via the Android app settings. Ideally, to follow EME recommendations, a persistent session is bound to an origin, only supported when first-party cookies are allowed, and cannot be re-opened when cookies are cleared.

\begin{table}[]
	\resizebox{\columnwidth}{!}{\begin{threeparttable}
	\begin{tabular}{|lll|c|c|c|c|c|}
		\hline
		\multicolumn{3}{|c|}{\textbf{Browsers}} & \multicolumn{2}{c|}{\textbf{RQ1}} & \multicolumn{2}{c|}{\textbf{RQ2}} & \multicolumn{1}{c|}{\textbf{RQ3}} \\ \hline
		\multicolumn{3}{|l|}{\textbf{Desktop}} & \multicolumn{1}{c|}{\textbf{VMP}} & \multicolumn{1}{c|}{\textbf{non-VMP}} & \multicolumn{1}{c|}{\textbf{VMP}} & \multicolumn{1}{c|}{\textbf{non-VMP}} &  \\\cline{4-7}
		& \multicolumn{2}{l|}{\textbf{Chromium Family}} &  &  &  &  & \\
		  &   & Chrome  &  $\newmoon$  & $\fullmoon$  & $\fullmoon$  & $\fullmoon$  & $\newmoon$ \\
		  &   & Edge    &  $\newmoon$  & $\fullmoon$  & $\fullmoon$  & $\fullmoon$  & $\newmoon$ \\
		  &   & Opera   &  $\newmoon$  & $\fullmoon$  & $\fullmoon$  & $\fullmoon$  & $\newmoon$ \\
		  &   & Brave   &  $\newmoon$  & $\fullmoon$  & $\fullmoon$  & $\fullmoon$  & N/A \\
		& \multicolumn{2}{l|}{\textbf{Firefox Family}} &  &  &  &  &  \\
		  &   & Firefox &  $\newmoon$  & $\fullmoon$  & $\fullmoon$  & $\fullmoon$  & N/A \\
		  &   & Tor     &  N/A  & N/A  & N/A  & N/A  & N/A \\
		  \hline
		\multicolumn{3}{|l|}{\textbf{ Android Mobile}} & \multicolumn{2}{l|}{} & \multicolumn{2}{l|}{} & \multicolumn{1}{l|}{} \\
		& \multicolumn{2}{l|}{\textbf{Chromium Family}} & \multicolumn{2}{l|}{} & \multicolumn{2}{l|}{} & \multicolumn{1}{l|}{} \\
		&  & Chrome & \multicolumn{2}{c|}{$\newmoon$} & \multicolumn{2}{c|}{$\newmoon$} & $\newmoon$ \\
		&  & Samsung & \multicolumn{2}{c|}{$\newmoon$} & \multicolumn{2}{c|}{$\newmoon$} & $\fullmoon$ \\
		&  & Edge & \multicolumn{2}{c|}{$\newmoon$} & \multicolumn{2}{c|}{$\newmoon$} & $\newmoon$ \\
		&  & Opera & \multicolumn{2}{c|}{$\newmoon$} & \multicolumn{2}{c|}{$\newmoon$} & $\fullmoon$ \\
		&  & Brave & \multicolumn{2}{c|}{N/A} & \multicolumn{2}{c|}{N/A} & N/A \\
		& \multicolumn{2}{l|}{\textbf{Firefox Family}} & \multicolumn{2}{l|}{} & \multicolumn{2}{l|}{} & \multicolumn{1}{l|}{} \\
		&  & Firefox  & \multicolumn{2}{c|}{$\fullmoon$} & \multicolumn{2}{c|}{$\fullmoon$} & N/A \\
		&  & Firefox Focus & \multicolumn{2}{c|}{$\fullmoon$} & \multicolumn{2}{c|}{$\fullmoon$} & N/A \\
		&  & Ghostery & \multicolumn{2}{c|}{$\fullmoon$} & \multicolumn{2}{c|}{$\fullmoon$} & N/A \\
		&  & Tor & \multicolumn{2}{c|}{N/A} & \multicolumn{2}{c|}{N/A} & N/A \\\hline
	\end{tabular}
		\begin{tablenotes}
			\small
			\item $\newmoon$ EME compliant.
			\item $\fullmoon$ Do not respect EME privacy recommendations. \\
		\end{tablenotes}
	\end{threeparttable}
	}
	\caption{Results for Implementation Questions per Browsers.}
	\label{table:prel_qr}
	\vspace{-1cm}
\end{table}

 \subsection{Results} \label{sec:impl_qr_results}
Here, we cover the findings obtained while following our methodology to answer our three research questions. A summary can be found in \autoref{table:prel_qr}. Tor and Brave mobile do not support EME, and therefore we were not able to perform any relevant experimentation on them. Thus, they are noted as N/A in our table.

\subsubsection{RQ1} 
As presented in \autoref{table:prel_qr}, we can see that the Privacy Mode and VMP are strongly bound to each other on desktop. Indeed, the Privacy Mode is enabled whenever the VMP is available on the underlying operating system. Thus, Privacy Mode is activated on Windows and macOS regardless of the browser used. In contrast, no browser protects the Client ID in license requests on Linux where VMP is not yet supported, as indicated by Widevine~\cite{widevine_oldnews}. We conclude that the Privacy Mode does only depend on the Widevine CDM that checks VMP before enforcement. No further action is required by browsers, explaining their identical behaviors.

As for mobiles, we find that browsers behave differently, and can be categorized into two families: Chromium and Firefox. The Chromium family includes Chrome (of course), Samsung Internet Browser, Edge, and Opera. As for the Firefox family, it includes Firefox, Firefox Focus, and Ghostery. Our experiments show the Client ID in clear for the Firefox family, implying no Privacy Mode. Only encrypted license requests were observed within the Chromium family. We push our analysis further to understand the rationale behind this divergence. Indeed, we reverse engineer browsers' apks with Jadx~\cite{jadx} to identify any calls to the \texttt{MediaDRM} class, which is the Android component communicating with the Widevine CDM. In particular, we spot the use of the \texttt{setPropertyString} method with the parameters \texttt{('PrivacyMode', 'enable')} for the Privacy Mode. Enforcing our claims, we encounter the call to this \texttt{setPropertyString} method only in the Chromium family. It is almost ironic that privacy-centered browsers, such as Firefox Focus and Ghostery, do not enable the Privacy Mode, thereby leaking sensitive and identifying data through the EME API.

\subsubsection{RQ2}
In this question, our methodology is to continue the EME workflow started previously as we force the event renewal following the license request generation in RQ1. Then, we investigate whether the Client ID is encrypted in the renewal request. Our evaluations complete our observations in RQ1. Indeed, we first observe that whenever the Client ID is not protected in the license request, it always appears in clear within renewal requests regardless of the execution environments. This concerns Linux systems on desktop and the Firefox family on mobile, and comes in accordance with the RQ1 findings. Second, we explore browsers on VMP-compatible systems. Once again, we notice that all browsers behave in the same way, strongly implying that the observed behavior is implemented by the CDM. Here, we witness the absence of encryption regarding the Client ID within the renewal request. Third, we analyze the mobile Chromium family. Our tests conclude that the Privacy Mode is also applied to the renewal event; the property set using the \texttt{MediaDRM} class enforces the Privacy Mode on all Client ID occurrences.

\subsubsection{RQ3}
Unlike the Privacy Mode, browsers differ greatly regarding persistent sessions. Recall that we study three properties: (1) preventing cross-origin sessions, (2) activation when blocking first-party cookies, and (3) availability after wiping cookies and site data. Here, we needed to exclude some browsers, since they do not provide any support for persistent sessions: the Firefox family on mobile as well as Firefox and Brave on desktop.

As for the examined properties, all browsers satisfy the first property, namely that sessions are only accessible by the origin, creating them from the same browsing profile. This implies that no malicious website can leverage sessions from other valid OTTs. Regarding the second property, we find that almost all browsers prevent persistent sessions when first-party cookies are blocked. The only exceptions are the mobile Opera and the Samsung Internet Browser. The same trend carries on for the third property. Indeed, only mobile Opera and the Samsung Internet Browser do not remove the persistent sessions after deleting cookies and site data, here note that erasing cookies cannot be done without erasing also site data from the browser settings. In fact, following our methodology, sessions can still be re-opened after wiping data for these two browsers. However, persistent sessions are indeed removed when clearing application data from the Android app settings. 

\subsection{Implications and Insights}
Nowadays, the EME standard is supported by almost all browsers~\cite{eme_browsers}, implying that any privacy leakage might impact a large user base, represented by our browser selection explained in \autoref{subsec:experimental_setup}. The W3C has sought to reassure the web community by minimizing such a risk. Tim Berners-Lee stated that ``\textit{the EME system can sandbox the DRM code to limit the damage it can do to the users privacy}''. In that sense, the EME specification includes various guidelines to enforce user privacy. Our experiments show that these guidelines are not technically enforced in practice. Indeed, the opaque nature of EME makes it hard to assess any privacy properties related to the use of a given proprietary module. More importantly, EME states that ``\textit{user-agents must take responsibility for providing users with adequate control over their own privacy}'', while allowing DRM providers to define their own format key contents that might involve distinguishing identifiers.

The EME specification (refer to section 11.4.2 of~\cite{EME}) presents their recommendations for a privacy-friendly EME implementation. Of particular interest, we mention two recommendations: (1) encrypt distinctive identifiers, and (2) treat key systems stored data like cookies or web storage. However, we argue that these considerations lack binding requirements and are insufficiently concrete. In that regard, users do not have any guarantee that their privacy is safeguarded when using EME with proprietary DRM modules. 

In our paper, we empirically challenge these privacy claims by scrutinizing one of the most popular real-world functioning of EME, namely Widevine. In our paper, we provide, to the best of our knowledge, the first conformance test verifying EME privacy recommendations for Widevine EME. Supported by our experimental setup and reverse engineering, our findings show a gap between these recommendations and deployed implementations of Widevine EME, implying a rather concerning scenario. 

First of all, we find that all desktop browsers fail to encrypt the Client ID, which represents distinctive characteristics in the Widevine protocol. Indeed, the Widevine CDM does not enable the Privacy Mode whenever VMP is not supported (e.g., Linux). Surprisingly, the CDM does not apply the Privacy Mode on renewal requests even when VMP is supported. On mobiles, privacy browsers do not encrypt the Client ID, although the Android platform provides all the necessary API for this purpose. Only the mobile Chromium family never reveals the Client ID in clear. As for the persistent sessions, the desktop browsers perform better, since the related data cannot be stored when the cookies are not allowed, and are always deleted when cookies are cleared. On mobile, the Chromium family acts differently. Chrome and Edge operate likewise their desktop counterparts. However, Opera and Samsung do not treat the stored data like cookies or web storage. For instance, there is no way to delete persistent sessions related data other than wiping all browser data from the Android app configurations. We notice that persistent sessions are not supported on all EME browsers: Firefox and Brave desktop do not offer such support. Overall, when supporting persistent sessions, browsers generally tend to comply with the EME specifications by managing related stored data as cookies. Otherwise, such data can be leveraged to track users having no easy mechanism to prevent their storage or even delete them.

\section{Privacy Implications}
\label{sec:privacy_considerations}

Our findings in \autoref{sec:ua_implem_privacy_mechanisms} point to a gap between EME and browsers in terms of privacy. In our analysis, we focus on the Client ID, aka distinctive identifier, and persistent sessions, namely session-related stored data. We continue our study to answer our second question: what is the privacy impact of this gap? Here, we refine our question to inspect device fingerprinting and user tracking. First, we examine whether the Client ID certificate chain constitutes a device fingerprint uniquely identifying the device starting EME. Second, we consider the Client Info attributes (refer to \autoref{tab:widevine_client_info}) and compare it with the User-Agent HTTP Header (UA) provided by browsers. Third, We discuss how persistent sessions can be leveraged for user tracking even when cookies are blocked or cleared. Below, we provide answers about each investigation. However, we start by reviewing the permissions required to perform the described actions; namely, we probe the permissions a website needs to call the EME API. We also present our experiment setup in a second time.

\subsection{Permissions} \label{subsec:permissions}
Our methodology is simple: we view some DRM-protected content on different browsers' fresh installation, and note whether users are asked to explicitly validate any related permission. Again, the Firefox and Chromium families behave differently. In the Chromium family, content protection service is allowed by default for all origins, leaving the user unaware of EME access by the web page. Regarding the Firefox family, permission is explicitly required for DRM access, but differs from desktop and mobile versions. On Android, permission is asked once per origin trying to access the EME API, without key systems distinction, while for desktop browsers, this permission is granted once for all origins and key systems. This implies that a desktop user granting DRM access to a legitimate OTT platform is also allowing illegitimate web pages to access EME. Firefox asks for permission by displaying the following message: Allow $<$URL$>$ to play DRM-controlled content? With two buttons, ``Don't allow'' and ``Allow'' with ``Allow'' pre-selected. 

\subsection{Experimental Setup} \label{subsec:experimental_setup}
To explore our question, we deployed our tests on multiple devices both for desktop and Android. Note that for both experiment setups, no real users were involved avoiding user privacy risks.

\noindent
\textbf{On Desktop.} For our population, we used 159,923 devices with Windows 10/11, macOS Big Sur and Linux. The majority comes from Windows and Mac VMs of BrowserStack\footnote{available on the platform \url{https://www.browserstack.com}}, added to our own device pool of Windows and Linux devices for comparison purpose between real and emulated devices. To inspect Widevine Client IDs and persistent sessions, we automated navigation to our custom site in which we open persistent sessions when available, and perform license renewal to obtain clear Client IDs using custom license policies as described in~\autoref{subsec:implem_methodo}. 

\noindent
\textbf{On Android.} We used 317 physical devices from BrowserStack and our own device pool, as well as 18,101 emulated ones, to collect Client IDs. We developed an Android app that communicates with the Widevine CDM, as a browser would do, to generate clear license requests and gather Client IDs in a fully automated way. For emulated Android devices, we used Google Pixel 1 to 6 from Android Studio~\cite{android_studio} with factory setting using Android version from 8.1 up to 13.

\noindent
\textbf{Emulated devices.} As mentioned in~\cite{DBLP:journals/popets/CasselLBWZBHJL22}, using emulated devices for web tracking measurement could have an impact on results relevance. Obviously, our created population cannot be used to represent a real world community; for instance, specific phone models and Android versions might be over-represented. However, we took great care to avoid that our results would be biased regarding the uniqueness of the collected identifiers. Indeed, the Client ID contains the RSA public key of the key pairs sent by the provisioning server. As pointed out in~\cite{DBLP:conf/dsn/PatatSF22}, the provisioning server generates a distinct Client ID for each Device ID. Provisioning with the same Device ID results in the same Client ID for the same app. Therefore, for license individualization on physical devices, Widevine defines a factory provisioning process to enforce the uniqueness of Device ID on Android devices. For our experimental settings, we built our population of Android emulated devices, so that each emulator has a unique Device ID, which is the default settings of the Android SDK tools. As for Desktop, the Client ID solely depends on the CDM version, regardless of the execution environment.

\subsection{Client ID Certificate Fingerprint} \label{subsec:clientif_cert_fingerprint}
In \autoref{subsec:widevine_client_id}, findings from our reverse engineering strongly indicate that the Client ID contains multiple distinctive permanent identifiers. Indeed, its inner Client Info contains device-specific attributes related to hardware instances and software versions for desktop and Android. In addition, it contains an RSA certificate chain used to protect the transmitted licenses. On the one hand, attributes can help classify devices into subgroups. On the other hand, the Widevine certificate chain might be enough to point out an exact device. The nature of this chain depends on the platform as explained above. On desktop (i.e., Linux, macOS, and Windows), the certificate chain is hard-coded within the CDM. Due to this design, all desktop Widevine CDMs share the same DRM certificate chain for a given CDM version. Thus, it does not enable fingerprinting on desktop. On Android devices, a Widevine-specific factory Root-of-Trust (RoT), called the \textit{Keybox}~\cite{WidevineSiteKeybox, DBLP:conf/sp/PatatSF22}, is used to perform a \textit{provisioning phase} requesting a certificate. This phase occurs once per app. The received certificate chain seems specific per device (or more precisely, Device ID). 

Below, we conduct more experiments to explore two important features: uniqueness and stability of the Client ID as a potential fingerprint. We only consider the certificate chain part of the Client ID.

\subsubsection{Uniqueness}
The purpose of a fingerprint is to completely characterize a specific device. Therefore, an important related feature is uniqueness: to what extent can we affirm that two license requests come from the same device if they contain the same chain? Here, we performed a clean installation and factory reset of devices for each test to evaluate certificate uniqueness in a \textit{multiple devices, single provisioning} approach using our EME website and custom Android app as explained in \autoref{subsec:experimental_setup}.

Our results are twofold. First, as expected, the Client ID is not unique on desktop environments; we only observed two different certificate serial numbers during our tests due to different CDM versions available between Windows and macOS/Linux. Second, the certificate chain is unique for Android devices. Indeed, We can precisely distinguish all mobile devices in the test set by extracting the certificate serial numbers of the gathered license requests. This implies that an Android device can be tracked by any web origin using the EME API when the Client ID is in clear, which is the case for all browsers among the Firefox family as shown in \autoref{sec:impl_qr_results}. We agree that our experiments do not constitute a definite proof of perfect uniqueness for an arbitrary device population, since this would have required a much larger test pool. Nevertheless, we argue that the technical details presented in~\cite{DBLP:conf/sp/PatatSF22} regarding provisioning make us believe that this fingerprint is unique by design. In fact, the Client ID contains the RSA public key of the key pairs sent by the provisioning server. To enforce individualization, the key pairs are not shared among other devices, confirming our findings.

\subsubsection{Stability} Recall that a unique fingerprint is not enough to track users. The leveraged fingerprint should resist changing with time in order to allow long-term tracking. For instance, a software-based fingerprint changes its value when the underlying software is updated, in contrast with a hardware-based one that is harder to collect. Here, we show that our fingerprint gets the best of two worlds: it is easy to collect (a few EME calls without explicit permissions) and stable over time. Here, we tested the stability of Client ID certificate with a \textit{single device, multiple provisioning} approach by removing the certificate instance to force new provisioning. On fresh settings, the CDM asks for certificate provisioning to create a license request. For desktop, forcing this behavior translates into removing the CDM library from the browser repository, while on Android, removing the stored certificate from the OS file system using privileged user rights, namely the \texttt{media} user level. We installed any CDM update when available, but we did not systematically attempt to install new applications for each provisioning. 

Similar to the uniqueness results, in desktop implementations, the certificate of the Client ID depends on the version of the CDM. During Widevine updates, the certificate serial number changes for all implementations, meaning stability on a version lifetime, which is 6 months in average. On Android, we adopted two longitudinal approaches: one device over two years, and multiple devices over one month. In both approaches, we removed the certificate to force triggering a new provisioning phase. In the former approach, we recovered the Client ID and removed the certificate at different intervals: one year, 6 months and 3 months. As for the latter, the deleting intervals are shorter: one week, one day, and one hour. After removal, we always get the same certificate, which strongly implies the stability of the Client ID.

\subsubsection{Comparative Analyses}
In order to put our results in perspective, we compare ourselves to three of the most large-scale studies of device fingerprinting on the web: Panopticlick~\cite{DBLP:conf/pet/Eckersley10}, AmIUnique~\cite{DBLP:conf/sp/LaperdrixRB16}, and Hiding in the Crowd~\cite{DBLP:conf/www/Gomez-BoixLB18}. These works inspect the uniqueness of fingerprints on mobile and desktop devices by gathering various attributes available through HTTP headers, plugins, and web APIs. Panopticlick, with 470,161 desktops, and AmIUnique, with 105,829 and 13,105 desktop and mobile devices, have shown a high proportion of uniqueness among gathered information with at least 81\% up to 94,2\% of unique fingerprints.

Later, Gómez-Boix et al. pointed out, with Hiding in the Crowd, that these results might be biased due to the tested population being privacy-aware users conscious of the collected values by the experiments. By fingerprinting 1,816,776 desktops and 251,166 mobiles, they found uniqueness of up to 35,7\% at best.

For our tests, we took the same order of magnitude for the number of devices as AmIUnique. Unlike AmIUnique, our experimental settings consist mainly of emulated devices. As explained previously, we took great attention to build our population, so that the Widevine provisioning ecosystem would process our requests similarly as it does to physical devices. While never unique for desktops, Client IDs of mobile devices offer a distinctive identifier with maximum uniqueness and stability through a JavaScript API available by any origin.

\subsection{Client Info as User-Agent Header}
As shown \autoref{subsec:widevine_client_id} and \autoref{fig:client_id}, the Client ID includes not only the fingerprint certificate chain, but also multiple other attributes within the Client Info structure. Here, we investigate the privacy leakage of the latter. The Client Info attributes include the CPU architecture, OS name, device name, and the complete device build name. Similar to the User-Agent HTTP Header (UA), these attributes can build a characteristic string for servers to identify the operating system, vendor, and version of the requesting browser or CDM. This is particularly interesting to get an augmented UA in which fields cannot be modified by some user-controlled browser configurations. Recall that the term \textit{user-agent} in EME refers to the software implementing the EME API (i.e., a browser). To avoid confusion in this subsection, we will use UA to refer to the browser User-Agent HTTP Header. 

Interestingly, the UA and the Client Info contain redundant information, such as the platform and architecture for both desktop and Android, or even the browser package certificate hash for mobile devices. Based on the UA structure~\cite{mozilla_useragent}, this redundant data covers most of the UA:

\begin{lstlisting}[mathescape=true]
	UA: 	Mozilla/5.0 ($\textbf{<system-information>}$) 
				<platform> (<platform-details>) 
				$\textbf{<extensions>}$
	e.g.: Mozilla/5.0 $\textbf{(Linux; Android 13; Pixel 4a)}$ 
				AppleWebKit/537.36 (KHTML, like Gecko) 
				$\textbf{Chrome/113.0.0.0 Mobile Safari/537.36}$
\end{lstlisting}

The resulted redundancy can be leveraged to detect browser instances attempting to hide or tamper with the UA in order to access specific resources (e.g., the mobile version of a website). This is possible since any UA modification, by spoofing for instance, has no impact on the Client Info attributes returned by Widevine. Therefore, Client Info can play the role of a never-lying UA. The reliability of Client Info is appealing for services relying on UA as a second factor authentication~\cite{DBLP:conf/uss/LinISP22}: lying users can be detected whenever the purported UA conflicts with the device class indicated by Client Info. Finally, minimizing the UA sent to origins is set as a goal for various privacy efforts, such as the Google Privacy Sandbox User-Agent Reduction proposal~\cite{google_ua}. The fact that some rich UA can be built from Widevine might limit the benefit of these current efforts.

\subsubsection{Uniqueness} As explained in \autoref{subsec:experimental_setup}, our dataset of mostly emulated devices cannot be relied upon to calculate the entropy of Client Info for a real-world population. Nevertheless, we can still provide some insights from our collected data. On Android, the data from Client Info can point to an exact device model and build version using the Build Info string (refer to the example of Pixel 7 in \autoref{tab:widevine_client_info}). On desktop, although with fewer fields than on Android, Client Info can still narrow down devices to architecture and OS subgroups, making entropy close to a standard browser UA regarding system information, which was studied in~\cite{DBLP:conf/www/Gomez-BoixLB18}. In the Chromium project, the related Widevine implementation indicates eight possible values\footnote{\url{https://chromium.googlesource.com/chromium/src/+/HEAD/third_party/widevine/cdm/widevine.gni\#23}}: Windows (x86, x64, and arm64), Linux (x64, arm, and arm64) and MacOS (x64, and arm64). It is important to note that Client Info is always obtained with the certificate chain of the Client ID, making its uniqueness less significant, especially on Android. 

\subsubsection{Stability} The Client Info stability depends mainly on the version of the Widevine CDM. On desktop, the available attributes are the OS and architecture being fixed. This leads to strong stability over time for desktop environment. On Android, the Client Info is influenced by the device build version that may change with the Widevine version, patch level, and build info, which makes the Client Info stability dependent on both Android and Widevine updates. The frequency of these updates depend heavily on mobile manufacturers; varying from monthly for Google Pixel phones and never for outdated devices no longer receiving any update. It is worth noting that stability is less relevant for Client Info when it is used to collect identifying data about devices, rather than fingerprinting.

\subsection{Persistent Session Tracking} 
Recall that all browsers from the Chromium family on Android and desktops support persistent sessions in Widevine, except for Brave. This support allows any web origin with EME API access to create a persistent session and collects its session ID for future usage. Persistent session IDs are random 16-byte identifiers represented as a hex string. When a client visits a website, it can be asked to open a previous session with persistent type using its ID; if the operation succeeds, this means that the client has already visited the website before. Thus, any website can check for the presence of a given session ID, revealing previous visitors. The approach is straightforward: attempt to open all the noted session IDs until a success occurs. This is possible because the number of failures is not limited. The scalability of this approach can be improved by breaking down the number of tested session IDs by device categorization. The privacy impact of re-invoking prior sessions is to achieve surreptitious cross-session tracking on the same origin even when cookies are blocked and site data are wiped (Android Opera and Samsung). For example, this allows an attacker to link user activity on the same site.

Based on our findings, persistent sessions are treated the same way as cookie data within browsers as mandated by EME, except for the mobile browsers Opera and Samsung Internet Browser. Oddly enough, they are not displayed in any Storage User Interface within navigator settings. We argue that this might mislead some users unaware of this mechanism storing cookie-like data. As a limitation compared to Client ID fingerprinting, this tracking mechanism is stateful. User tracking through persistent sessions requires the origin to keep track of session IDs due to the lack of API that lists stored sessions within the browser file system. 
 
\section{EME User Tracking Scripts}

In sections \ref{sec:ua_implem_privacy_mechanisms} and \ref{sec:privacy_considerations}, we discussed how these EME misimplementations could be exploited to track users either by a stateful cookie-like token or by a unique fingerprint with strong stability. In this section, we detail how a curious website could use these mechanisms to track users over the Internet based on our threat model described in \autoref{sec:introduction}. In particular, they set renewal request intervals, force Client ID within EME renewal messages, or even open persistent sessions. To do so, we developed a tool in  JavaScript code that sends the Client ID of a user-agent to a malicious server and opens persistent sessions when supported. The script is available on our GitHub~\footnote{\url{https://github.com/Avalonswanderer/widevine_eme_fingerprinting}} for reproducibility purposes. More importantly, our tool can test the conformance of implementations with EME privacy concerns. In the following, we present our workflow depicted in \autoref{fig:fingerprint_script_flow}.

\subsection{EME Track}
\begin{figure}
	\centering
	\resizebox{\columnwidth}{!}{\includegraphics[scale=1.175]{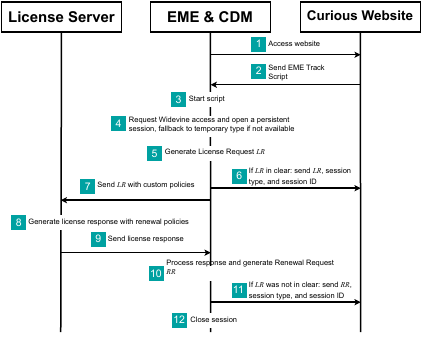}
	}
	\caption{EME Widevine Fingerprint Flow.}
	\label{fig:fingerprint_script_flow}
	\vspace{-2em}
\end{figure} Like our experiment settings, we leverage the Widevine staging license server with custom Widevine policies support to operate. 

From \autoref{fig:fingerprint_script_flow}, in steps~\framebox(7,7){\small 1} and~\framebox(7,7){\small 2}, a user visits a curious origin and receives the EME Track script. The user-agent starts executing it~\framebox(7,7){\small 3}, and is asked to instantiate its underlying Widevine CDM with a default configuration with persistent license type support. If not supported, temporary type is used for sessions~\framebox(7,7){\small 4}. Here, we assume that users enable EME-related permissions (according to \autoref{subsec:permissions}).

In step~\framebox(7,7){\small 5}, initialization data are provided to Widevine through the \texttt{generateRequest} API. These data are related to Widevine pre-encrypted media, available with the staging license server. When the user-agent is in Privacy Mode by default, a service certificate is required and provided by the script to the CDM. Here, this certificate corresponds to the \texttt{license.widevine.com} domain. Once processed by the CDM, the initialization data are used to create a license request, noted \textit{LR}. If \textit{LR} is in clear, we forward it in~\framebox(7,7){\small 6}, with its session ID and type, to the curious website. The workflow continues from here to complete a standard license provisioning by sending \textit{LR} to the Widevine license server. While doing so, custom Widevine policies are used (see~\autoref{subsec:widevine_license_policy}), requiring persistent rights, renewable license, and renewal attempts within a few seconds for the demanded license.

Steps~\framebox(7,7){\small 8} and~\framebox(7,7){\small 9} correspond to the license server response with appropriate content key(s) and rights. The user-agent provides the response to the CDM in step~\framebox(8,8){\small 10}, allowing the session to be saved on the file system if persistent. Within a few seconds, license policies force the Widevine CDM to generate a renewal request \textit{RR} including the Client ID of the device. If the previous \textit{LR} Client ID was not in clear text, the user-agent sends \textit{RR}, session ID, and session type to the curious site in step~\framebox(8,8){\small 11} before closing the session in~\framebox(8,8){\small 12}.

With either \textit{LR} or \textit{RR} in clear text, a malicious website can gather unique fingerprints with device-specific attributes for mobile devices and a never-lying UA for desktop implementations. With the support of persistent licenses, the received session ID can be used to track users if requests are not distinctive, as shown below.

\subsection{Persistent Session Re-invocation}
With our EME Track script, an origin can gather Client IDs in clear when available from any user-agent and create persistent sessions when supported within the browser file system. Here we describe the workflow of a curious website tracking users based on their stored Widevine persistent sessions.

At first, a user-agent visits a curious website and receives a tracking script starting to be executed. A Widevine access is demanded using \texttt{requestMediaKeySystemAccess} with persistent license support configuration. At this point, either by hard-coded values within the script or by receiving it through the origin, previously gathered session IDs are provided by the curious website to the CDM in an attempt to \texttt{load} an already existing session. On return, the script provides the result of this operation to the curious website. In the event of user-agent revisiting the website, a persistent session can be opened, allowing the origin to identify the user.

\section{Discussion}

\subsection{Privacy Leakage in Practice}
The adoption of EME by W3C has been a controversial decision~\cite{DBLP:conf/space/Halpin17}. Much of the raised debate was related to the opaque design of EME in which no technical guarantees can be given about the security and privacy properties of an EME-compliant system. It is true that, technically, like any software, EME has the potential to be privacy-invasive and to have possible security issues. However, we have shown that the controversy is well-founded due to the privacy concerns inherent in uniquely identifying keys in Widevine.

The answers to our two high-level questions are: (1) yes, there is a gap between the promised guarantees by EME and the actual Widevine implementations, (2) and this leaks private information, allowing long-term user tracking. Indeed, the Client ID is not appropriately protected and can be recovered in most browsers. This Client ID allows not only building an augmented UA on which browsers have no control, but also a stable and unique fingerprint on mobile environments. Moreover, EME might store some persistent offline data, that can be invoked subsequently to track visitors. All browsers mislead users by not displaying these EME data in any Storage User Interface. Some browsers do not even include them when wiping site data. More importantly, we show how easy it is to leverage EME to leak data, especially since browsers mostly adopt calling EME ``\textit{without annoying and confusing consent prompts}''~\cite{w3c_archives}.

\subsection{OTTs as EME Actors}
Section~\ref{sec:eme_wv} and~\autoref{table:content_messages} show the content of EME messages when used with Widevine. During our tests, we observed EME workflow from various legitimate OTTs: Amazon Prime Video, Netflix, and Disney+. We found out that in addition to these arguments, Netflix and Disney+ were using an OTT-specific field within license response and renewal request/response. OTTs can leverage their own license servers to use this field and store information within the client CDM as they please to, for instance, manage communication between client and server statelessly. However, this particular usage changes the EME compliance chain by introducing another actor: OTTs themselves. With such a field, another opaque layer is added to EME messages increasing privacy risks in the event of identifying data. The complexity of involving OTTs chosen data within EME-specific messages is illustrated with Netflix in~\autoref{appendix:netflix_case}.

\subsection{License Server as Big Brother} \label{subsec:big_brother}
The EME standard specifies that CDM ``\textit{implementations should avoid use of distinctive identifiers}'', otherwise, ``\textit{the CDM vendor may be able to track the activity of the user}''. Despite this recommendation, Widevine defines and shares the Client ID, which encompasses some fingerprinting properties. As explained in \autoref{subsec:privacy_vmp}, Widevine attempts to limit the damage by introducing the Privacy Mode in which, if enabled, the Client ID gets encrypted with a public key provided by the license server. Recall that OTTs can provide their own Widevine-signed server certificate using \texttt{setServerCertificate}. Consequently, these OTTs can get the Client ID of any device, with considerable privacy implications on Android mobiles due to a unique and stable fingerprint. 

Some might argue that sharing such a distinctive identifier is vital for DRM functioning to bound licenses to some individualization process. However, this violates users' privacy, given the fact that all of this is done without users' consent in an opaque way. We are concerned that Widevine and OTTs, via their license servers, are able to obtain identifying information on any device. Indeed, EME states that ``user-agents must take responsibility for providing users with adequate control over their own privacy''. Our paper shows that such a statement about EME respecting user privacy is misleading and is contradicted by the real-world operating of Widevine. Browsers have no choice but to share the Client ID with license servers whenever EME is enabled, breaking privacy in numerous contexts. We appeal that it becomes compelling for proprietary DRM systems to take responsibility and transparently enforce all EME design suggestions, especially for privacy and security.

\subsection{Responsible Disclosure}
Our findings have been timely communicated to all concerned parties following responsible disclosure processes. Mozilla Firefox was quite responsive, and we got rewarded via their bug bounty program. The Mozilla EME team investigated our findings and released a patch to address the identified privacy issues and acknowledged us in the Mozilla Hall of Fame. Regarding Client ID being in clear in renewal requests, we first contacted the EME Chrome team that reviewed our disclosure report and showed concerns about its privacy consequence, namely the EME user-agent. They confirmed our intuition that the problem is caused by the Widevine CDM. Therefore, we filed a Widevine bug report about missing Privacy Mode on VMP systems, and are still in communication with them. 
\section{Related Work}

\subsection{Reverse-Engineering Widevine}
Widevine provides a proprietary DRM solution. Despite its widespread, there is not much literature studying Widevine security. This lack of public analysis is due to the DMCA 1201 clause that makes it illegal to study DRM systems. Fortunately, since 2018, security researchers can freely investigate and publish security flaws when acting with ``good intentions''. Patat et al.~\cite{DBLP:conf/sp/PatatSF22} explored the undocumented Widevine protocol and detailed its different cryptographic components on Android. Zhao~\cite{blackhat_zhao} investigated and broke the TEE-based Android Widevine. Both works give deep insights on Widevine design, but its W3C EME integration was left unstudied. In this paper, we inspect the EME messages when Widevine-protected content is being played over a browser to investigate privacy implication.

\subsection{Privacy and W3C APIs}
One of the missions of W3C is to design and standardize new APIs to implement new features over the web. In this mission, W3C strives for interoperability between service providers operating with different message formats. Thus, the APIs are designed to work smoothly with industrial-controlled messages that might contain identifying information breaking privacy. The W3C acknowledges this risk, highlighting the necessity to create privacy-safe features.

Certain W3C APIs have already been abused to break privacy. Indeed, Olejnik et al.~\cite{DBLP:conf/esorics/OlejnikACD15} investigated the privacy impact of the HTML5 Battery Status API defined by W3C~\cite{battery_api}. They observed a possible fingerprint by collecting the battery charge level and charging timing. Moreover, Doty et al.~\cite{DBLP:journals/corr/abs-1003-1775} studied the W3C Geolocation API used to transfer the device location information to the origin in an opaque way and found possible abusive usage. Our work takes a step in this direction, and provides insights into privacy leakage caused by opaque EME implementations despite all the privacy concerns raised in the recommendations.

\subsection{Fingerprinting through DRM Systems}
Our work does not constitute the first work leveraging DRM for device 
fingerprinting. Indeed, in 2020, the FingerprintJs Android project~\cite{fingerprintjs_android} (written in Kotlin) was released to collect numerous device identifiers that are available to an Android app, including a Widevine identifier called \textit{Device ID}. This identifier is accessible through the \texttt{MediaDRM} API, and is unique per device~\cite{DBLP:conf/sp/PatatSF22}, allowing device fingerprinting. Our work differs from this project in two ways. First, the used Device ID of Widevine is uniquely accessible from the Android \texttt{MediaDRM} API with \texttt{getPropertyByteArray}, a method not reachable from the browser JS APIs, thereby requiring higher privileges (e.g., app-level instead of browser enclaves). Second, this ID is not linked to the certification chain but is directly included in the Widevine Keybox RoT only available on Android. Therefore, this technique does not extend to desktop.

In 2020, the Reddit website asked for DRM access to list all the available key systems proposed by the user-agent~\cite{fingerprint_reddit_drm}. This method can be used to add additional attributes to other fingerprinting values. Nevertheless, these attributes are common to many devices, rarely narrowing down user-agents to individual groups. %
 
\section{Conclusion}

During its standardization, EME was decried because no effective control can be performed on the underlying closed-source DRM systems to check whether they comply both in terms of security and privacy. In this paper, we show that these privacy concerns are founded by investigating the closed-source Widevine DRM system through the EME API. We present how a curious origin can leverage the Widevine protocol and misimplementation within browsers to gather unique and with long-term stability fingerprints for android mobile devices and a never-lying User-Agent header for both mobile and desktop. In addition to this stateless tracking, we exploit a stateful mechanism that websites can misuse to track users online. We provide insights on current user-agent implementations' compliance with EME privacy guidelines. We hope this work will encourage further research on user-agent privacy compliance and provide arguments against opaque systems within web standards.

\begin{acks}
We thank the anonymous reviewers and our shepherd for their valuable comments, helping us to improve this document. This work benefited from the support of the project ANR-22-CE39-0005 DRAMA of the French National Research Agency (ANR).
\end{acks}

\bibliographystyle{ACM-Reference-Format}
\bibliography{ref}


\begin{thebibliography}{67}


\ifx \showCODEN    \undefined \def \showCODEN     #1{\unskip}     \fi
\ifx \showDOI      \undefined \def \showDOI       #1{#1}\fi
\ifx \showISBNx    \undefined \def \showISBNx     #1{\unskip}     \fi
\ifx \showISBNxiii \undefined \def \showISBNxiii  #1{\unskip}     \fi
\ifx \showISSN     \undefined \def \showISSN      #1{\unskip}     \fi
\ifx \showLCCN     \undefined \def \showLCCN      #1{\unskip}     \fi
\ifx \shownote     \undefined \def \shownote      #1{#1}          \fi
\ifx \showarticletitle \undefined \def \showarticletitle #1{#1}   \fi
\ifx \showURL      \undefined \def \showURL       {\relax}        \fi
\providecommand\bibfield[2]{#2}
\providecommand\bibinfo[2]{#2}
\providecommand\natexlab[1]{#1}
\providecommand\showeprint[2][]{arXiv:#2}

\bibitem[Acar et~al\mbox{.}(2014)]%
        {DBLP:conf/ccs/AcarEEJND14}
\bibfield{author}{\bibinfo{person}{Gunes Acar}, \bibinfo{person}{Christian
  Eubank}, \bibinfo{person}{Steven Englehardt}, \bibinfo{person}{Marc
  Ju{\'{a}}rez}, \bibinfo{person}{Arvind Narayanan}, {and}
  \bibinfo{person}{Claudia D{\'{\i}}az}.} \bibinfo{year}{2014}\natexlab{}.
\newblock \showarticletitle{The Web Never Forgets: Persistent Tracking
  Mechanisms in the Wild}. In \bibinfo{booktitle}{\emph{Proceedings of the 2014
  {ACM} {SIGSAC} Conference on Computer and Communications Security,
  Scottsdale, AZ, USA, November 3-7, 2014}},
  \bibfield{editor}{\bibinfo{person}{Gail{-}Joon Ahn}, \bibinfo{person}{Moti
  Yung}, {and} \bibinfo{person}{Ninghui Li}} (Eds.).
  \bibinfo{publisher}{{ACM}}, \bibinfo{pages}{674--689}.
\newblock
\urldef\tempurl%
\url{https://doi.org/10.1145/2660267.2660347}
\showDOI{\tempurl}


\bibitem[{Android}(2023)]%
        {android_studio}
\bibfield{author}{\bibinfo{person}{{Android}}.}
  \bibinfo{year}{2023}\natexlab{}.
\newblock \bibinfo{title}{{Android Studio}}.
\newblock \bibinfo{howpublished}{\url{https://developer.android.com/studio/}}.
\newblock


\bibitem[{Apple}(2023)]%
        {fairplay}
\bibfield{author}{\bibinfo{person}{{Apple}}.} \bibinfo{year}{2023}\natexlab{}.
\newblock \bibinfo{title}{{Apple FairPlay}}.
\newblock
  \bibinfo{howpublished}{\url{https://developer.apple.com/streaming/fps/}}.
\newblock


\bibitem[Ayenson et~al\mbox{.}(2011)]%
        {Ayenson2011FlashCA}
\bibfield{author}{\bibinfo{person}{Mika~D. Ayenson},
  \bibinfo{person}{Dietrich~James Wambach}, \bibinfo{person}{Ashkan Soltani},
  \bibinfo{person}{Nathaniel Good}, {and} \bibinfo{person}{Chris~Jay
  Hoofnagle}.} \bibinfo{year}{2011}\natexlab{}.
\newblock \showarticletitle{Flash Cookies and Privacy II: Now with HTML5 and
  ETag Respawning}.
\newblock


\bibitem[Berners-Lee(2017)]%
        {eme_debate_1}
\bibfield{author}{\bibinfo{person}{Tim Berners-Lee}.}
  \bibinfo{year}{2017}\natexlab{}.
\newblock \bibinfo{title}{{On EME in HTML5}}.
\newblock
  \bibinfo{howpublished}{\url{https://www.w3.org/blog/2017/02/on-eme-in-html5}}.
\newblock


\bibitem[Blog(2022)]%
        {netflix_esn}
\bibfield{author}{\bibinfo{person}{Netflix~Technology Blog}.}
  \bibinfo{year}{2022}\natexlab{}.
\newblock \bibinfo{title}{{Modernizing the Netflix TV UI Deployment Process}}.
\newblock
  \bibinfo{howpublished}{\url{https://netflixtechblog.medium.com/modernizing-the-netflix-tv-ui-deployment-process-28e022edaaef}}.
\newblock


\bibitem[Brave(2022)]%
        {brave50Mactiveusers}
\bibfield{author}{\bibinfo{person}{Brave}.} \bibinfo{year}{2022}\natexlab{}.
\newblock \bibinfo{title}{{Brave Passes 50 Million Monthly Active Users,
  Growing 2x for the Fifth Year in a Row}}.
\newblock \bibinfo{howpublished}{\url{https://brave.com/2021-recap/}}.
\newblock


\bibitem[{Brave Software, Inc.}(2022)]%
        {brave_browser}
\bibfield{author}{\bibinfo{person}{{Brave Software, Inc.}}}
  \bibinfo{year}{2022}\natexlab{}.
\newblock \bibinfo{title}{{Brave Browser}}.
\newblock \bibinfo{howpublished}{\url{https://brave.com/}}.
\newblock


\bibitem[{Can I Use}(2023)]%
        {eme_browsers}
\bibfield{author}{\bibinfo{person}{{Can I Use}}.}
  \bibinfo{year}{2023}\natexlab{}.
\newblock \bibinfo{title}{{Encrypted Media Extensions}}.
\newblock \bibinfo{howpublished}{\url{https://caniuse.com/eme}}.
\newblock


\bibitem[Cassel et~al\mbox{.}(2022)]%
        {DBLP:journals/popets/CasselLBWZBHJL22}
\bibfield{author}{\bibinfo{person}{Darion Cassel}, \bibinfo{person}{Su{-}Chin
  Lin}, \bibinfo{person}{Alessio Buraggina}, \bibinfo{person}{William Wang},
  \bibinfo{person}{Andrew Zhang}, \bibinfo{person}{Lujo Bauer},
  \bibinfo{person}{Hsu{-}Chun Hsiao}, \bibinfo{person}{Limin Jia}, {and}
  \bibinfo{person}{Timothy Libert}.} \bibinfo{year}{2022}\natexlab{}.
\newblock \showarticletitle{OmniCrawl: Comprehensive Measurement of Web
  Tracking With Real Desktop and Mobile Browsers}.
\newblock \bibinfo{journal}{\emph{Proc. Priv. Enhancing Technol.}}
  \bibinfo{volume}{2022}, \bibinfo{number}{1} (\bibinfo{year}{2022}),
  \bibinfo{pages}{227--252}.
\newblock


\bibitem[Dorwin et~al\mbox{.}(2019)]%
        {EME}
\bibfield{author}{\bibinfo{person}{David Dorwin}, \bibinfo{person}{Jerry
  Smith}, \bibinfo{person}{Mark Watson}, {and} \bibinfo{person}{Adrian
  Bateman}.} \bibinfo{year}{2019}\natexlab{}.
\newblock \bibinfo{title}{{Encrypted Media Extensions}}.
\newblock \bibinfo{howpublished}{\url{https://www.w3.org/TR/encrypted-media/}}.
\newblock


\bibitem[Doty et~al\mbox{.}(2010)]%
        {DBLP:journals/corr/abs-1003-1775}
\bibfield{author}{\bibinfo{person}{Nick Doty}, \bibinfo{person}{Deirdre~K.
  Mulligan}, {and} \bibinfo{person}{Erik Wilde}.}
  \bibinfo{year}{2010}\natexlab{}.
\newblock \showarticletitle{Privacy Issues of the {W3C} Geolocation {API}}.
\newblock \bibinfo{journal}{\emph{CoRR}}  \bibinfo{volume}{abs/1003.1775}
  (\bibinfo{year}{2010}).
\newblock
\showeprint[arXiv]{1003.1775}
\urldef\tempurl%
\url{http://arxiv.org/abs/1003.1775}
\showURL{%
\tempurl}


\bibitem[Eckersley(2010)]%
        {DBLP:conf/pet/Eckersley10}
\bibfield{author}{\bibinfo{person}{Peter Eckersley}.}
  \bibinfo{year}{2010}\natexlab{}.
\newblock \showarticletitle{How Unique Is Your Web Browser?}. In
  \bibinfo{booktitle}{\emph{Privacy Enhancing Technologies, 10th International
  Symposium, {PETS} 2010, Berlin, Germany, July 21-23, 2010. Proceedings}}
  \emph{(\bibinfo{series}{Lecture Notes in Computer Science},
  Vol.~\bibinfo{volume}{6205})}, \bibfield{editor}{\bibinfo{person}{Mikhail~J.
  Atallah} {and} \bibinfo{person}{Nicholas~J. Hopper}} (Eds.).
  \bibinfo{publisher}{Springer}, \bibinfo{pages}{1--18}.
\newblock
\urldef\tempurl%
\url{https://doi.org/10.1007/978-3-642-14527-8\_1}
\showDOI{\tempurl}


\bibitem[Fingerprintjs(2020)]%
        {fingerprintjs_android}
\bibfield{author}{\bibinfo{person}{Fingerprintjs}.}
  \bibinfo{year}{2020}\natexlab{}.
\newblock \bibinfo{title}{{FingerprintJS Android}}.
\newblock
  \bibinfo{howpublished}{\url{https://github.com/fingerprintjs/fingerprintjs-android}}.
\newblock


\bibitem[Forum(2022)]%
        {dash_drm_id}
\bibfield{author}{\bibinfo{person}{Dash~Industry Forum}.}
  \bibinfo{year}{2022}\natexlab{}.
\newblock \bibinfo{title}{{Content Protection}}.
\newblock
  \bibinfo{howpublished}{\url{https://dashif.org/identifiers/content_protection/}}.
\newblock


\bibitem[Foundation(2014)]%
        {eme_fsf_condemns}
\bibfield{author}{\bibinfo{person}{Free~Software Foundation}.}
  \bibinfo{year}{2014}\natexlab{}.
\newblock \bibinfo{title}{{FSF condemns partnership between Mozilla and Adobe
  to support Digital Restrictions Management}}.
\newblock
  \bibinfo{howpublished}{\url{https://www.fsf.org/news/fsf-condemns-partnership-between-mozilla-and-adobe-to-support-digital-restrictions-management}}.
\newblock


\bibitem[Gizmodo(2018)]%
        {safari_fp}
\bibfield{author}{\bibinfo{person}{Gizmodo}.} \bibinfo{year}{2018}\natexlab{}.
\newblock \bibinfo{title}{{Apple Declares War on 'Browser Fingerprinting,' the
  Sneaky Tactic That Tracks You in Incognito Mode}}.
\newblock
  \bibinfo{howpublished}{\url{https://gizmodo.com/apple-declares-war-on-browser-fingerprinting-the-sneak-1826549108}}.
\newblock


\bibitem[G{\'{o}}mez{-}Boix et~al\mbox{.}(2018)]%
        {DBLP:conf/www/Gomez-BoixLB18}
\bibfield{author}{\bibinfo{person}{Alejandro G{\'{o}}mez{-}Boix},
  \bibinfo{person}{Pierre Laperdrix}, {and} \bibinfo{person}{Benoit Baudry}.}
  \bibinfo{year}{2018}\natexlab{}.
\newblock \showarticletitle{Hiding in the Crowd: an Analysis of the
  Effectiveness of Browser Fingerprinting at Large Scale}. In
  \bibinfo{booktitle}{\emph{Proceedings of the 2018 World Wide Web Conference
  on World Wide Web, {WWW} 2018, Lyon, France, April 23-27, 2018}},
  \bibfield{editor}{\bibinfo{person}{Pierre{-}Antoine Champin},
  \bibinfo{person}{Fabien Gandon}, \bibinfo{person}{Mounia Lalmas}, {and}
  \bibinfo{person}{Panagiotis~G. Ipeirotis}} (Eds.).
  \bibinfo{publisher}{{ACM}}, \bibinfo{pages}{309--318}.
\newblock
\urldef\tempurl%
\url{https://doi.org/10.1145/3178876.3186097}
\showDOI{\tempurl}


\bibitem[Google(2022a)]%
        {google_privacysandbox}
\bibfield{author}{\bibinfo{person}{Google}.} \bibinfo{year}{2022}\natexlab{a}.
\newblock \bibinfo{title}{{The Privacy Sandbox}}.
\newblock \bibinfo{howpublished}{\url{https://privacysandbox.com/}}.
\newblock


\bibitem[Google(2022b)]%
        {google_ua}
\bibfield{author}{\bibinfo{person}{Google}.} \bibinfo{year}{2022}\natexlab{b}.
\newblock \bibinfo{title}{{User-Agent reduction}}.
\newblock
  \bibinfo{howpublished}{\url{https://developer.chrome.com/docs/privacy-sandbox/user-agent/}}.
\newblock


\bibitem[Google(2023a)]%
        {firefoxfocus_gp}
\bibfield{author}{\bibinfo{person}{Google}.} \bibinfo{year}{2023}\natexlab{a}.
\newblock \bibinfo{title}{{Google Play Store: Firefox Focus: No Fuss Browser}}.
\newblock
  \bibinfo{howpublished}{\url{https://play.google.com/store/apps/details?id=org.mozilla.focus}}.
\newblock


\bibitem[Google(2023b)]%
        {ghostery_gp}
\bibfield{author}{\bibinfo{person}{Google}.} \bibinfo{year}{2023}\natexlab{b}.
\newblock \bibinfo{title}{{Google Play Store: Ghostery Privacy Browser}}.
\newblock
  \bibinfo{howpublished}{\url{https://play.google.com/store/apps/details?id=com.ghostery.android.ghostery}}.
\newblock


\bibitem[{Google Widevine}(2021)]%
        {widevine_oldnews}
\bibfield{author}{\bibinfo{person}{{Google Widevine}}.}
  \bibinfo{year}{2021}\natexlab{}.
\newblock \bibinfo{title}{Widevine {N}ews}.
\newblock
  \bibinfo{howpublished}{\url{https://web.archive.org/web/20210903150435/https://www.widevine.com/news}}.
\newblock


\bibitem[{Google Widevine}(2023)]%
        {widevineSite}
\bibfield{author}{\bibinfo{person}{{Google Widevine}}.}
  \bibinfo{year}{2023}\natexlab{}.
\newblock \bibinfo{title}{{Widevine}}.
\newblock \bibinfo{howpublished}{\url{https://widevine.com/}}.
\newblock


\bibitem[Hadad(2021)]%
        {tomer_widevinel3decryptor_wiki}
\bibfield{author}{\bibinfo{person}{Tomer Hadad}.}
  \bibinfo{year}{2021}\natexlab{}.
\newblock \bibinfo{title}{{Reversing the old Widevine Content Decryption
  Module}}.
\newblock
  \bibinfo{howpublished}{\url{https://github.com/tomer8007/widevine-l3-decryptor/wiki/Reversing-the-old-Widevine-Content-Decryption-Module}}.
\newblock


\bibitem[Halpin(2017)]%
        {DBLP:conf/space/Halpin17}
\bibfield{author}{\bibinfo{person}{Harry Halpin}.}
  \bibinfo{year}{2017}\natexlab{}.
\newblock \showarticletitle{{The Crisis of Standardizing {DRM:} The Case of
  {W3C} Encrypted Media Extensions}}. In \bibinfo{booktitle}{\emph{{SPACE}}}
  \emph{(\bibinfo{series}{Lecture Notes in Computer Science},
  Vol.~\bibinfo{volume}{10662})}. \bibinfo{publisher}{Springer},
  \bibinfo{pages}{10--29}.
\newblock


\bibitem[H\'egaret(2017)]%
        {eme_critics}
\bibfield{author}{\bibinfo{person}{Philippe~Le H\'egaret}.}
  \bibinfo{year}{2017}\natexlab{}.
\newblock \bibinfo{title}{{Disposition of Comments for Encrypted Media
  Extensions and Director's decision}}.
\newblock
  \bibinfo{howpublished}{\url{https://lists.w3.org/Archives/Public/public-html-media/2017Jul/0000.html}}.
\newblock


\bibitem[Insights(2022)]%
        {video_streaming_money}
\bibfield{author}{\bibinfo{person}{Fortune~Business Insights}.}
  \bibinfo{year}{2022}\natexlab{}.
\newblock \bibinfo{title}{{Video Streaming Market Size}}.
\newblock
  \bibinfo{howpublished}{\url{https://www.fortunebusinessinsights.com/video-streaming-market-103057}}.
\newblock


\bibitem[{ISO Central Secretary}(2016)]%
        {iso_cenc_2016}
\bibfield{author}{\bibinfo{person}{{ISO Central Secretary}}.}
  \bibinfo{year}{2016}\natexlab{}.
\newblock \bibinfo{booktitle}{\emph{{Information technology — MPEG systems
  technologies — Part 7: Common encryption in ISO base media file format
  files}}}.
\newblock \bibinfo{type}{Standard} {ISO/IEC 23001-7:2016}.
  \bibinfo{institution}{International Organization for Standardization}.
\newblock
\urldef\tempurl%
\url{https://www.iso.org/standard/68042.html}
\showURL{%
\tempurl}


\bibitem[{ISO Central Secretary}(2022)]%
        {iso_dash_2022}
\bibfield{author}{\bibinfo{person}{{ISO Central Secretary}}.}
  \bibinfo{year}{2022}\natexlab{}.
\newblock \bibinfo{booktitle}{\emph{{Information technology — Dynamic
  adaptive streaming over HTTP (DASH) — Part 1: Media presentation
  description and segment formats}}}.
\newblock \bibinfo{type}{Standard} {ISO/IEC 23009-1:2022}.
  \bibinfo{institution}{International Organization for Standardization}.
\newblock
\urldef\tempurl%
\url{https://www.iso.org/standard/83314.html}
\showURL{%
\tempurl}


\bibitem[Jaffe(2017)]%
        {eme_debate_2}
\bibfield{author}{\bibinfo{person}{Jeff Jaffe}.}
  \bibinfo{year}{2017}\natexlab{}.
\newblock \bibinfo{title}{{Reflections on the EME debate}}.
\newblock
  \bibinfo{howpublished}{\url{https://www.w3.org/blog/2017/09/reflections-on-the-eme-debate}}.
\newblock


\bibitem[Kamkar(2010)]%
        {evercookie_samykamkar}
\bibfield{author}{\bibinfo{person}{Samy Kamkar}.}
  \bibinfo{year}{2010}\natexlab{}.
\newblock \bibinfo{title}{{evercookie}}.
\newblock \bibinfo{howpublished}{\url{https://samy.pl/evercookie/}}.
\newblock


\bibitem[kkapsner(2022)]%
        {canvasblocker}
\bibfield{author}{\bibinfo{person}{kkapsner}.} \bibinfo{year}{2022}\natexlab{}.
\newblock \bibinfo{title}{{CanvasBlocker}}.
\newblock
  \bibinfo{howpublished}{\url{https://addons.mozilla.org/en-US/firefox/addon/canvasblocker/}}.
\newblock


\bibitem[Kostiainen et~al\mbox{.}(2022)]%
        {battery_api}
\bibfield{author}{\bibinfo{person}{Anssi Kostiainen}, \bibinfo{person}{Mounir
  Lamouri}, {and} \bibinfo{person}{Raphael~Kubo da Costa}.}
  \bibinfo{year}{2022}\natexlab{}.
\newblock \bibinfo{title}{{Battery Status API}}.
\newblock \bibinfo{howpublished}{\url{https://www.w3.org/TR/battery-status/}}.
\newblock


\bibitem[Laperdrix et~al\mbox{.}(2016)]%
        {DBLP:conf/sp/LaperdrixRB16}
\bibfield{author}{\bibinfo{person}{Pierre Laperdrix}, \bibinfo{person}{Walter
  Rudametkin}, {and} \bibinfo{person}{Benoit Baudry}.}
  \bibinfo{year}{2016}\natexlab{}.
\newblock \showarticletitle{Beauty and the Beast: Diverting Modern Web Browsers
  to Build Unique Browser Fingerprints}. In \bibinfo{booktitle}{\emph{{IEEE}
  Symposium on Security and Privacy, {SP} 2016, San Jose, CA, USA, May 22-26,
  2016}}. \bibinfo{publisher}{{IEEE} Computer Society},
  \bibinfo{pages}{878--894}.
\newblock
\urldef\tempurl%
\url{https://doi.org/10.1109/SP.2016.57}
\showDOI{\tempurl}


\bibitem[Lin et~al\mbox{.}(2022)]%
        {DBLP:conf/uss/LinISP22}
\bibfield{author}{\bibinfo{person}{Xu Lin}, \bibinfo{person}{Panagiotis Ilia},
  \bibinfo{person}{Saumya Solanki}, {and} \bibinfo{person}{Jason Polakis}.}
  \bibinfo{year}{2022}\natexlab{}.
\newblock \showarticletitle{Phish in Sheep's Clothing: Exploring the
  Authentication Pitfalls of Browser Fingerprinting}. In
  \bibinfo{booktitle}{\emph{{USENIX} Security Symposium}}.
  \bibinfo{publisher}{{USENIX} Association}, \bibinfo{pages}{1651--1668}.
\newblock


\bibitem[Maone(2022)]%
        {noscript}
\bibfield{author}{\bibinfo{person}{Giorgio Maone}.}
  \bibinfo{year}{2022}\natexlab{}.
\newblock \bibinfo{title}{{NoScript Official Website}}.
\newblock \bibinfo{howpublished}{\url{https://noscript.net/}}.
\newblock


\bibitem[Mayer(2009)]%
        {mayer2009any}
\bibfield{author}{\bibinfo{person}{Jonathan~R Mayer}.}
  \bibinfo{year}{2009}\natexlab{}.
\newblock \showarticletitle{{Any person... a pamphleteer: Internet Anonymity in
  the Age of Web 2.0}}.
\newblock \bibinfo{journal}{\emph{Undergraduate Senior Thesis, Princeton
  University}}  \bibinfo{volume}{85} (\bibinfo{year}{2009}).
\newblock


\bibitem[{Mdn Web Docs}(2023)]%
        {mozilla_useragent}
\bibfield{author}{\bibinfo{person}{{Mdn Web Docs}}.}
  \bibinfo{year}{2023}\natexlab{}.
\newblock \bibinfo{title}{{User-Agent}}.
\newblock
  \bibinfo{howpublished}{\url{https://developer.mozilla.org/en-US/docs/Web/HTTP/Headers/User-Agent
  }}.
\newblock


\bibitem[{Microsoft}(2023)]%
        {playready}
\bibfield{author}{\bibinfo{person}{{Microsoft}}.}
  \bibinfo{year}{2023}\natexlab{}.
\newblock \bibinfo{title}{{Microsoft PlayReady}}.
\newblock \bibinfo{howpublished}{\url{https://www.microsoft.com/playready/}}.
\newblock


\bibitem[Mowery et~al\mbox{.}(2011)]%
        {MBYS11}
\bibfield{author}{\bibinfo{person}{Keaton Mowery}, \bibinfo{person}{Dillon
  Bogenreif}, \bibinfo{person}{Scott Yilek}, {and} \bibinfo{person}{Hovav
  Shacham}.} \bibinfo{year}{2011}\natexlab{}.
\newblock \showarticletitle{Fingerprinting Information in {JavaScript}
  Implementations}. In \bibinfo{booktitle}{\emph{Proceedings of W2SP 2011}},
  \bibfield{editor}{\bibinfo{person}{Helen Wang}} (Ed.). IEEE Computer Society.
\newblock


\bibitem[Mowery and Shacham(2012)]%
        {MS12}
\bibfield{author}{\bibinfo{person}{Keaton Mowery} {and} \bibinfo{person}{Hovav
  Shacham}.} \bibinfo{year}{2012}\natexlab{}.
\newblock \showarticletitle{Pixel Perfect: Fingerprinting Canvas in {HTML5}}.
  In \bibinfo{booktitle}{\emph{Proceedings of W2SP 2012}},
  \bibfield{editor}{\bibinfo{person}{Matt Fredrikson}} (Ed.). IEEE Computer
  Society.
\newblock


\bibitem[Mozilla(2023)]%
        {firefox_protection}
\bibfield{author}{\bibinfo{person}{Mozilla}.} \bibinfo{year}{2023}\natexlab{}.
\newblock \bibinfo{title}{{Enhanced Tracking Protection in Firefox for
  desktop}}.
\newblock
  \bibinfo{howpublished}{\url{https://support.mozilla.org/en-US/kb/enhanced-tracking-protection-firefox-desktop}}.
\newblock


\bibitem[Mulazzani et~al\mbox{.}(2013)]%
        {mulazzani2013fast}
\bibfield{author}{\bibinfo{person}{Martin Mulazzani}, \bibinfo{person}{Philipp
  Reschl}, \bibinfo{person}{Markus Huber}, \bibinfo{person}{Manuel Leithner},
  \bibinfo{person}{Sebastian Schrittwieser}, \bibinfo{person}{Edgar Weippl},
  {and} \bibinfo{person}{FC Wien}.} \bibinfo{year}{2013}\natexlab{}.
\newblock \showarticletitle{Fast and reliable browser identification with
  javascript engine fingerprinting}. In \bibinfo{booktitle}{\emph{Web 2.0
  Workshop on Security and Privacy (W2SP)}}, Vol.~\bibinfo{volume}{5}.
  Citeseer, \bibinfo{pages}{4}.
\newblock


\bibitem[Multilogin(2022)]%
        {canvasdefender}
\bibfield{author}{\bibinfo{person}{Multilogin}.}
  \bibinfo{year}{2022}\natexlab{}.
\newblock \bibinfo{title}{{Canvas Defender}}.
\newblock
  \bibinfo{howpublished}{\url{https://addons.mozilla.org/en-US/firefox/addon/no-canvas-fingerprinting/}}.
\newblock


\bibitem[Netflix(2020)]%
        {netflix_msl}
\bibfield{author}{\bibinfo{person}{Netflix}.} \bibinfo{year}{2020}\natexlab{}.
\newblock \bibinfo{title}{{Message Security Layer}}.
\newblock \bibinfo{howpublished}{\url{https://github.com/Netflix/msl}}.
\newblock


\bibitem[Olejnik et~al\mbox{.}(2015)]%
        {DBLP:conf/esorics/OlejnikACD15}
\bibfield{author}{\bibinfo{person}{Lukasz Olejnik}, \bibinfo{person}{Gunes
  Acar}, \bibinfo{person}{Claude Castelluccia}, {and} \bibinfo{person}{Claudia
  D{\'{\i}}az}.} \bibinfo{year}{2015}\natexlab{}.
\newblock \showarticletitle{The Leaking Battery - {A} Privacy Analysis of the
  {HTML5} Battery Status {API}}. In \bibinfo{booktitle}{\emph{Data Privacy
  Management, and Security Assurance - 10th International Workshop, {DPM} 2015,
  and 4th International Workshop, {QASA} 2015, Vienna, Austria, September
  21-22, 2015. Revised Selected Papers}} \emph{(\bibinfo{series}{Lecture Notes
  in Computer Science}, Vol.~\bibinfo{volume}{9481})},
  \bibfield{editor}{\bibinfo{person}{Joaqu{\'{\i}}n Garc{\'{\i}}a{-}Alfaro},
  \bibinfo{person}{Guillermo Navarro{-}Arribas}, \bibinfo{person}{Alessandro
  Aldini}, \bibinfo{person}{Fabio Martinelli}, {and} \bibinfo{person}{Neeraj
  Suri}} (Eds.). \bibinfo{publisher}{Springer}, \bibinfo{pages}{254--263}.
\newblock
\urldef\tempurl%
\url{https://doi.org/10.1007/978-3-319-29883-2\_18}
\showDOI{\tempurl}


\bibitem[Patat et~al\mbox{.}(2022a)]%
        {DBLP:conf/sp/PatatSF22}
\bibfield{author}{\bibinfo{person}{Gwendal Patat}, \bibinfo{person}{Mohamed
  Sabt}, {and} \bibinfo{person}{Pierre{-}Alain Fouque}.}
  \bibinfo{year}{2022}\natexlab{a}.
\newblock \showarticletitle{Exploring Widevine for Fun and Profit}. In
  \bibinfo{booktitle}{\emph{43rd {IEEE} Security and Privacy, {SP} Workshops
  2022, San Francisco, CA, USA, May 22-26, 2022}}. \bibinfo{publisher}{{IEEE}},
  \bibinfo{pages}{277--288}.
\newblock
\urldef\tempurl%
\url{https://doi.org/10.1109/SPW54247.2022.9833867}
\showDOI{\tempurl}


\bibitem[Patat et~al\mbox{.}(2022b)]%
        {DBLP:conf/dsn/PatatSF22}
\bibfield{author}{\bibinfo{person}{Gwendal Patat}, \bibinfo{person}{Mohamed
  Sabt}, {and} \bibinfo{person}{Pierre{-}Alain Fouque}.}
  \bibinfo{year}{2022}\natexlab{b}.
\newblock \showarticletitle{WideLeak: How Over-the-Top Platforms Fail in
  Android}. In \bibinfo{booktitle}{\emph{52nd Annual {IEEE/IFIP} International
  Conference on Dependable Systems and Networks, {DSN} 2022, Baltimore, MD,
  USA, June 27-30, 2022}}. \bibinfo{publisher}{{IEEE}},
  \bibinfo{pages}{501--508}.
\newblock
\urldef\tempurl%
\url{https://doi.org/10.1109/DSN53405.2022.00056}
\showDOI{\tempurl}


\bibitem[Project(2022)]%
        {tor_browser}
\bibfield{author}{\bibinfo{person}{Tor Project}.}
  \bibinfo{year}{2022}\natexlab{}.
\newblock \bibinfo{title}{{Tor Browser}}.
\newblock \bibinfo{howpublished}{\url{https://www.torproject.org/}}.
\newblock


\bibitem[rlaphoenix(2022)]%
        {pywidevine}
\bibfield{author}{\bibinfo{person}{rlaphoenix}.}
  \bibinfo{year}{2022}\natexlab{}.
\newblock \bibinfo{title}{{pywidevine}}.
\newblock
  \bibinfo{howpublished}{\url{https://github.com/rlaphoenix/pywidevine}}.
\newblock


\bibitem[Schwartz(2001)]%
        {schwartz_2001}
\bibfield{author}{\bibinfo{person}{John Schwartz}.}
  \bibinfo{year}{2001}\natexlab{}.
\newblock \bibinfo{title}{{Giving Web a Memory Cost Its Users Privacy}}.
\newblock
  \bibinfo{howpublished}{\url{https://www.nytimes.com/2001/09/04/business/giving-web-a-memory-cost-its-users-privacy.html}}.
\newblock


\bibitem[Skylot(2019)]%
        {jadx}
\bibfield{author}{\bibinfo{person}{Skylot}.} \bibinfo{year}{2019}\natexlab{}.
\newblock \bibinfo{title}{{Jadx - Dex to Java decompiler}}.
\newblock \bibinfo{howpublished}{\url{https://github.com/skylot/jadx}}.
\newblock


\bibitem[Smitop(2020)]%
        {fingerprint_reddit_drm}
\bibfield{author}{\bibinfo{person}{Smitop}.} \bibinfo{year}{2020}\natexlab{}.
\newblock \bibinfo{title}{{Reddit's website uses DRM for fingerprinting}}.
\newblock \bibinfo{howpublished}{\url{https://iter.ca/post/reddit-whiteops/}}.
\newblock


\bibitem[Soltani et~al\mbox{.}(2010)]%
        {DBLP:conf/aaaiss/SoltaniCMTH10}
\bibfield{author}{\bibinfo{person}{Ashkan Soltani}, \bibinfo{person}{Shannon
  Canty}, \bibinfo{person}{Quentin Mayo}, \bibinfo{person}{Lauren Thomas},
  {and} \bibinfo{person}{Chris~Jay Hoofnagle}.}
  \bibinfo{year}{2010}\natexlab{}.
\newblock \showarticletitle{Flash Cookies and Privacy}. In
  \bibinfo{booktitle}{\emph{Intelligent Information Privacy Management, Papers
  from the 2010 {AAAI} Spring Symposium, Technical Report SS-10-05, Stanford,
  California, USA, March 22-24, 2010}}. \bibinfo{publisher}{{AAAI}}.
\newblock
\urldef\tempurl%
\url{http://www.aaai.org/ocs/index.php/SSS/SSS10/paper/view/1070}
\showURL{%
\tempurl}


\bibitem[{S}tat {C}ounter(2022a)]%
        {browserDesktopMarketShare}
\bibfield{author}{\bibinfo{person}{{S}tat {C}ounter}.}
  \bibinfo{year}{2022}\natexlab{a}.
\newblock \bibinfo{title}{{Desktop Browser Market Share Worldwide}}.
\newblock
  \bibinfo{howpublished}{\url{https://gs.statcounter.com/browser-market-share/desktop/worldwide/2022}}.
\newblock


\bibitem[{S}tat {C}ounter(2022b)]%
        {browserMobileMarketShare}
\bibfield{author}{\bibinfo{person}{{S}tat {C}ounter}.}
  \bibinfo{year}{2022}\natexlab{b}.
\newblock \bibinfo{title}{{Mobile Browser Market Share Worldwide}}.
\newblock
  \bibinfo{howpublished}{\url{https://gs.statcounter.com/browser-market-share/mobile/worldwide/2022}}.
\newblock


\bibitem[Tech Emmy~Awards(2018)]%
        {eme_emmy_award}
\bibfield{author}{\bibinfo{person}{The~Insider Tech Emmy~Awards}.}
  \bibinfo{year}{2018}\natexlab{}.
\newblock \bibinfo{title}{{70th Award Recipients}}.
\newblock
  \bibinfo{howpublished}{\url{https://theemmys.tv/tech-70th-award-recipients}}.
\newblock


\bibitem[{Tor Project}(2023)]%
        {tor2022users}
\bibfield{author}{\bibinfo{person}{{Tor Project}}.}
  \bibinfo{year}{2023}\natexlab{}.
\newblock \bibinfo{title}{{Tor Metrics User Stats 2022}}.
\newblock
  \bibinfo{howpublished}{\url{https://metrics.torproject.org/userstats-relay-country.html?start=2022-01-01&end=2023-01-01&country=all&events=off}}.
\newblock


\bibitem[W3C(2017)]%
        {eme_recommendations}
\bibfield{author}{\bibinfo{person}{W3C}.} \bibinfo{year}{2017}\natexlab{}.
\newblock \bibinfo{title}{{W3C Publishes Encrypted Media Extensions (EME) as a
  W3C Recommendation}}.
\newblock
  \bibinfo{howpublished}{\url{https://www.w3.org/2017/09/pressrelease-eme-recommendation.html.en}}.
\newblock


\bibitem[Watson(2017a)]%
        {w3c_archives}
\bibfield{author}{\bibinfo{person}{Mark Watson}.}
  \bibinfo{year}{2017}\natexlab{a}.
\newblock \bibinfo{title}{{Response from Director to formal objection ``Turn
  off EME by default and activate only with express permission from user''}}.
\newblock
  \bibinfo{howpublished}{\url{https://lists.w3.org/Archives/Public/public-html-media/2017Apr/0013.html}}.
\newblock


\bibitem[Watson(2017b)]%
        {web_crypto_api}
\bibfield{author}{\bibinfo{person}{Mark Watson}.}
  \bibinfo{year}{2017}\natexlab{b}.
\newblock \bibinfo{title}{{Web Cryptography API}}.
\newblock \bibinfo{howpublished}{\url{https://www.w3.org/TR/WebCryptoAPI}}.
\newblock


\bibitem[West and Pulimood(2012)]%
        {10.5555/2038772.2038791}
\bibfield{author}{\bibinfo{person}{William West} {and}
  \bibinfo{person}{S.~Monisha Pulimood}.} \bibinfo{year}{2012}\natexlab{}.
\newblock \showarticletitle{Analysis of Privacy and Security in HTML5 Web
  Storage}.
\newblock \bibinfo{journal}{\emph{J. Comput. Sci. Coll.}} \bibinfo{volume}{27},
  \bibinfo{number}{3} (\bibinfo{date}{jan} \bibinfo{year}{2012}),
  \bibinfo{pages}{80–87}.
\newblock
\showISSN{1937-4771}


\bibitem[{Widevine}(2023a)]%
        {WidevineSiteKeybox}
\bibfield{author}{\bibinfo{person}{{Widevine}}.}
  \bibinfo{year}{2023}\natexlab{a}.
\newblock \bibinfo{title}{Widevine {DRM}}.
\newblock
  \bibinfo{howpublished}{\url{https://www.widevine.com/solutions/widevine-drm}}.
\newblock


\bibitem[{Widevine}(2023b)]%
        {widevine_integration_platform}
\bibfield{author}{\bibinfo{person}{{Widevine}}.}
  \bibinfo{year}{2023}\natexlab{b}.
\newblock \bibinfo{title}{{Widevine Integration Platform Shaka Player Tool}}.
\newblock
  \bibinfo{howpublished}{\url{https://integration.widevine.com/player}}.
\newblock


\bibitem[zackmark29(2022)]%
        {NetflixESNGenerator}
\bibfield{author}{\bibinfo{person}{zackmark29}.}
  \bibinfo{year}{2022}\natexlab{}.
\newblock \bibinfo{title}{{NetflixESNGenerator}}.
\newblock
  \bibinfo{howpublished}{\url{https://github.com/zackmark29/NetflixESNGenerator}}.
\newblock


\bibitem[Zhao(2021)]%
        {blackhat_zhao}
\bibfield{author}{\bibinfo{person}{Qi Zhao}.} \bibinfo{year}{2021}\natexlab{}.
\newblock \bibinfo{title}{Wideshears: Investigating and breaking widevine on
  QTEE}.
\newblock \bibinfo{howpublished}{BlackHat Asia}.
\newblock


\end{thebibliography}

\newpage
\appendix
\section{Widevine License Policy Fields} \label{appendix:widevine_license_policy}
\begin{table}[h]
	\resizebox{\columnwidth}{!}{\begin{tabular}{cll}
		\hline
		\hline
		\textit{Code} & \textit{Policy Name}              & \textit{Description}                                                            \\
		\hline
		\rowcolor{gray!20}
		0x08                & Can Play                          & Allow playback.                                                            \\
		0x10                & Can Persist                       & \Gape[0pt][2pt]{\makecell[l]{Allow the license to be\\stored for offline usage.}} \\
		\rowcolor{gray!20}
		0x18                & Can Renew                         & \Gape[0pt][2pt]{\makecell[l]{License can be renewed\\to extend duration.}} \\
		0x20                & Rental Duration Seconds           & \Gape[0pt][2pt]{\makecell[l]{Time in seconds for which\\playback is allowed.}} \\
		\rowcolor{gray!20}
		0x28                & Playback Duration Seconds         & \Gape[0pt][2pt]{\makecell[l]{Time in seconds for which\\playback is allowed once started.}} \\
		0x30                & License Duration Seconds          & \Gape[0pt][2pt]{\makecell[l]{Time in seconds for which\\a license is valid.}} \\
		\rowcolor{gray!20}
		0x38                & \Gape[0pt][2pt]{\makecell[l]{Renewal Recovery\\Duration Seconds}} & \Gape[0pt][2pt]{\makecell[l]{Time in seconds for which\\playback is allowed until\\successful license renew.}} \\
		0x42                & Renewal Server URL                & \Gape[0pt][2pt]{\makecell[l]{URL of the license server\\for renewal request.}} \\
		\rowcolor{gray!20}
		0x48                & Renewal Delay Seconds             & \Gape[0pt][2pt]{\makecell[l]{Time in seconds before\\trying to renew a license,\\starting from its usage.}} \\
		0x50                & \Gape[0pt][2pt]{\makecell[l]{Renewal Retry\\Interval Seconds}} & \Gape[0pt][2pt]{\makecell[l]{Time in seconds between\\renewal requests\\if renew is unsuccessful.}} \\
		\rowcolor{gray!20}
		0x58                & Renew with Usage                  & \Gape[0pt][2pt]{\makecell[l]{Try to renew license\\ as soon as it is used.}}  \\
		0x60                & Always include Client ID          & \Gape[0pt][2pt]{\makecell[l]{Force the Client ID presence\\within renewal request.}} \\
		\rowcolor{gray!20}
		0x70                & \Gape[0pt][2pt]{\makecell[l]{Soft Enforce Playback\\Duration}}    & \Gape[0pt][2pt]{\makecell[l]{Enforce playback duration\\based on license duration.}} \\
		0x78                & \Gape[0pt][2pt]{\makecell[l]{Soft Enforce Rental\\Duration}}      & \Gape[0pt][2pt]{\makecell[l]{Enforce rental duration\\based on license duration.}} \\
		\rowcolor{gray!20}
		0x80                & Watermarking Control              & Specify a watermarking system. \\
		\hline
		\hline
	\end{tabular}
	}
\label{table:widevine_policy}
\end{table} 
\section{Detailed Web Browser Versions} \label{appendix:browser_versions}
\begin{table}[h]
	\resizebox{\columnwidth}{!}{\begin{tabular}{ll}
		\hline
		\hline
		\textit{Category}                         & \textit{Browser}                                      \\
		\hline
		\rowcolor{gray!20}
		                                 & Chrome 109.0.5114.120                   \\
		\rowcolor{gray!20}
		                                 & Edge 109.0.1518.70                      \\
		\rowcolor{gray!20}
		                                 & Opera 94.0.4606.76                      \\
		\rowcolor{gray!20}
		                                 & Firefox 108.0.2                         \\
		\rowcolor{gray!20}
		                                 & Brave 1.47.186, Chromium 109.0.5114.119 \\
		\rowcolor{gray!20}
		\multirow{-6}{*}{Desktop Browsers} & Tor 12.0                                \\
										  & Chrome 109.0.5114.85                    \\
		                                 & Edge 109.0.1518.70                      \\
		                                 & Samsung Internet Browser 19.0.6.3               \\
		                                 & Opera 73.1.3844.69816                   \\
		                                 & Firefox 109.1.1                         \\
		                                 & Firefox Focus 108.1.0                   \\
		                                 & Brave 1.47.175, Chromium 109.0.5414.87  \\
		                                 & Tor 102.2.1                             \\
		\multirow{-9}{*}{Mobile Browsers} & Ghostery Browser build \#2015907851 \\ \hline \hline                 
	\end{tabular}
	}
\label{table:browser_version}
\end{table}
 
\section{The Case of Netflix} \label{appendix:netflix_case}

When considering the EME messages generated by Netflix, we observe the usage by Netflix of an OTT-specific field in the license response, renewal request, and renewal response. Our experiments show that this field includes the required Movie ID and the device ESN (Equipment Serial Number) defined by Netflix. According to~\cite{netflix_esn}, the Netflix ESN (NESN) is used in production to send updates to a specific pool of devices to beta-test new fixes and features. This NESN is generated by appending several data~\cite{NetflixESNGenerator}: device category (e.g., Chrome browser, aka ChromeCDM for all desktops, Smartphones, Tablets, Android TVs, Smart Displays, Google TVs), the version of the OEM Crypto Library, the manufacturer and the model of the device followed by a random string that is constant for a given device. Therefore, similar to the Client ID, the NESN should be encrypted to protect users' privacy in case a proxy is used while streaming.

Thus, we extend our research questions and investigate NESN leakage: (1) is it encrypted? and (2) how does the NESN constitute a fingerprint? While focusing on Netflix, we consider the same experiment settings as described in \autoref{subsec:re_settings}. It is worth noting that, in license response, Netflix transmits licenses that require renewal a few seconds later. Therefore, we were able to easily observe the whole EME workflow without particular policies from the user-agents context. For both desktop and mobile, from the point of view of EME, the NESN is never protected on license and renewal responses. Nevertheless, Netflix EME messages containing this NESN are always encrypted through the Message Security Layer (MSL) protocol, leveraging Widevine as a cryptographic library, therefore not relying solely on HTTPS. As a matter of fact, the Web Cryptography API~\cite{web_crypto_api} on browsers has been specifically standardized by Netflix within the W3C for this purpose~\cite{netflix_msl}. We conclude that Netflix attempts to keep the NESN private by masking EME messages over the network. On mobile, Netflix only runs with its own application, with no browser support; however, the presence of this NESN within EME-specific messages in desktop implementations involves the OTT in the protocol consideration for privacy. The MSL being out of the EME recommendation, this encryption does not comply with privacy concerns by adding the NESN as a distinctive identifier within opaque messages.

Concerning the second question, we examine the privacy loss from a possible NESN leakage. As often in our study, the fallout depends on whether the leakage occurs on mobile or desktop. For mobile, the NESN ends with a 64-byte string. This string is unique per device for a given Widevine security level (i.e., L1 and L3). In addition, we note that the related fingerprint is stable since it is preserved even after wiping all app data using the Android settings. For desktop, the ending string is composed of 30 bytes. Unlike the Client ID, the NESN is unique per device, thereby might constituting a fingerprint if leaked through EME before MSL usage. Fortunately, the desktop NESN value is linked to cookies and updated after cookies and site data deletion, falling back to standard cookie concerns. We performed our tests on Windows, macOS, and Linux. 

Our study of Netflix shows that the privacy guarantees of EME are quite brittle. Indeed, it depends on the implementations of three parties that are proprietary and reluctant to communicate: web browsers, DRM systems, and streaming services. The Netflix case is an example of how complex the streaming ecosystem is. Netflix defines an identifying ESN and argues that it is necessary to beta test its new features, thereby compliant with GDPR. However, at the same time, this same ESN brings a distinctive identifier, within the EME protocol, under the responsibility of the OTT.

\end{document}